\documentclass[twocolumn,dvipsnames]{aastex63}

\pdfoutput=1 %for arXiv submission
\usepackage{amsmath,amstext}
\usepackage[T1]{fontenc}
\usepackage[figure,figure*]{hypcap}
\usepackage{graphicx,epsf}

\usepackage{amsmath}
\usepackage{amssymb}
\usepackage{braket}
\usepackage{amssymb}
\usepackage{color}
\usepackage{xcolor}
\usepackage{graphicx}
\usepackage{latexsym}
\usepackage{listings}
\usepackage{mathrsfs}
\usepackage{letltxmacro}
\usepackage[normalem]{ulem}

\usepackage{subfigure}
\usepackage{verbatim}
\usepackage{tabularx}
\usepackage{ragged2e}
\usepackage{multirow}
\usepackage{amsbsy}

\usepackage{numprint}
\usepackage{makecell}
\usepackage[utf8]{inputenc}

\newcommand{\R}{\mathbb{R}}
\newcommand{\N}{\mathbb{N}} % Integers
\newcommand{\V}{\mathcal{V}}

\newcommand{\F}{\mathcal{F}}

\newcommand{\M}{\mathcal{M}}
\newcommand{\I}{\mathcal{I}}

\newcommand{\B}{\text{B}}

\npdecimalsign{.}
\npthousandsep{,}

\newcolumntype{Y}{>{\RaggedRight\arraybackslash}X}

\graphicspath{{./figures/}}

\sloppypar

\shorttitle{MESA Modeling of Interacting Binaries}
\shortauthors{Hernandez et al.}
\graphicspath{{./}{figures/}}
\begin{document}

\title{Simulations of Interacting Binary Systems - Pathways to Radio Bright GRB Progenitors}

\correspondingauthor{}
\email{aherna82@depaul.edu}

\author[0000-0003-3850-2498]{Angel Hernandez}
\altaffiliation{McNair Scholar, Science Undergraduate Laboratory Internships (SULI) Fellow}
\affiliation{Department of Physics, University of Colorado at Boulder;  Boulder, CO 80309-0390, USA}
\affiliation{Department of Physics and Astrophysics, DePaul University;  Chicago, IL 60614, USA}

\affiliation{CCS-2, Computational Physics and Methods, Los Alamos National Laboratory, Los Alamos, NM 87544, USA}

\author[0000-0002-4854-8636]{Roseanne M.~Cheng}
\affiliation{CCS-2, Computational Physics and Methods, Los Alamos National Laboratory, Los Alamos, NM 87544, USA}
\affiliation{T-3, Fluid Dynamics and Solid Mechanics, Los Alamos National Laboratory, Los Alamos, NM 87544, USA}
\affiliation{Department of Physics and Astronomy, University of New Mexico, Albuquerque, NM 87106, USA}
\affiliation{Space Science Institute,
4765 Walnut St, Suite B,
Boulder, CO 80301}
\author[0000-0003-1707-7998]{Nicole M. Lloyd-Ronning}
\affiliation{CCS-2, Computational Physics and Methods, Los Alamos National Laboratory, Los Alamos, NM 87544, USA}
\affiliation{Center for Theoretical Astrophysics, Los Alamos National Laboratory, Los Alamos, NM 87544, USA}
\affiliation{Department of Math, Science and Engineering, University of New Mexico, Los Alamos, NM 87544, USA}

\author[0000-0002-8925-057X]{C.~E.~Fields}
\altaffiliation{Feynman Distinguished Postdoctoral Fellow}
\affiliation{CCS-2, Computational Physics and Methods, Los Alamos National Laboratory, Los Alamos, NM 87544, USA}
\affiliation{Center for Theoretical Astrophysics, Los Alamos National Laboratory, Los Alamos, NM 87544, USA}
%\affiliation{Computer, Computational, and Statistical Sciences Division, Los Alamos National Laboratory, Los Alamos, NM 87545, USA}
\affiliation{Steward Observatory, University of Arizona, Tucson, AZ 85721, USA}

\begin{abstract}
    Although the association of gamma-ray bursts with massive stellar death is on firm footing, the nature of the progenitor system and the key ingredients required for a massive star to produce a gamma-ray burst remain open questions.  {Here, we investigate the evolution of a $15-25M_\odot$ massive star with a $10-15 M_\odot$ black hole using \textsc{MESA}.}  {We quantify companion influenced angular momentum evolution over stellar lifetime for orbital periods where tides are significant, varying stellar and black hole masses, initial stellar spin, and accretion and dynamo prescriptions while tracking mass loss and angular momentum.} {Final spin is set by tidal torques versus stellar winds.} {For binaries that initially avoid Roche lobe overflow, tides can spin up the star, but late stage expansion can drive tidal stripping; associated mass and angular momentum loss can suppress spin up.}  {We find that massive star black hole binaries at comparable mass ratios may be potential GRB progenitors for short orbital periods ($\sim 20 - 5\times10^2$ days) and long orbital periods ($\sim 2\times10^3 - 4\times10^3$ days), where our suite of lifetime simulations reveals a favored parameter space with negligible mass loss and enough spin angular momentum to power a GRB jet.}  {For initially non rotating stars, this provides a lower limit on final spin above a threshold estimate consistent with forming a post collapse black hole mass of $5-10M_\odot$ with spin parameter $\geq 0.5$.}  {For initially rapidly rotating stars, tidal interactions may sustain high spin when mass loss is negligible because the binary is not tidally synchronized.}
\end{abstract}

\keywords{Gamma-ray bursts --- Black Holes --- Interacting Binaries }

\section{Introduction}

Gamma-ray bursts (GRBs) are the most luminous objects in the universe. The connection between so-called long GRBs (lGRBs; those with prompt gamma-ray emission lasting more than two seconds) and the deaths of massive stars  (MS) is strongly supported, both theoretically and observationally, including the direct association of supernova events with many lGRBs \citep{Woos93,MW99, BKD02, Hjorth03,WB06,WH06,KNJ08a, KNJ08b, HB12, Ly17}.  The idea is that, upon collapse, a rapidly spinning MS will form a rapidly spinning compact-object central engine (usually assumed to be a black hole (BH) with a surrounding accretion disk, although neutron star central engines are also possible) that is capable of launching a relativistic jet due to a strong build-up of Poynting along the spin axis of the BH through the so-called Blandford-Znajek process \citep{BZ77}. While stars with high angular momentum and a stripped hydrogen envelope are necessary ingredients for a successful jet launch, the exact conditions required to produce an lGRB from a collapsing star \citep{MW99,YL05,HMM05,Yoon06,WH06}, and indeed the jet launching mechanism itself  \cite[e.g.][]{BK08, KB09, LR19bz,KP21}, remain uncertain.  It is an ongoing pursuit to determine and understand {\em which} stars make GRBs, and {\em why}.

  This mystery has been further deepened by the recent evidence \citep{LRF17, LR19, Chak23} that those GRBs with radio afterglows (so-called radio-bright GRBs) appear to have significantly {\em longer-lasting prompt gamma-ray duration} and higher isotropic equivalent energy than their radio-dark counterparts.  It has been suggested \citep{LR22} that these radio bright GRBs may be a result of a MS collapsing in a system with an influential compact object companion. Under the right conditions, the companion can spin-up the MS, providing it the angular momentum it needs to create a longer lasting jet and therefore longer-lasting prompt GRB; this companion may also cause a more extended and dense circumburst environment, providing the surrounding gas needed for the GRB to shine brightly in the radio band.  

 In this paper, we examine the influence of a BH companion on a MS over a range of phase space varying orbital period, metallicity, initial spin of the star, and { stellar and BH masses} and discuss the implications this has for the final angular momentum of the star and the potential GRB it may produce.  We identify a particular region of the binary parameter space where the MS is subject to negligible mass loss from both wind and Roche-lobe overflow (RLOF) as the tidal effects significantly increase the spin angular momentum prior to stellar collapse at the end of its evolution.  We use the well-developed, open-source stellar evolution code: \emph{Modules for Experiments in Stellar Astrophysics} (\textsc{MESA}) to perform this study.  We present the parameter space in which this system will produce a highly-spinning MS.

 The systems we consider, with these particular mass ratios, have been observed in a number of X-ray binary systems \citep{Kelley83, Lev93, Lev00, Tv06, Fal15}, and recently \cite{Klen25} presented a detailed study of mass transfer in MS binary systems, showing that these types of systems will most likely have orbital separations around $r \gtrsim 10R_{\odot}$.  However, a presentation of how the BH companion might affect the end-state of the MS in these systems, in the context of being a progenitor for a GRB, to our knowledge has not been done.  Preliminary population synthesis models of these systems \citep{Cas24} and (A. Cason et al. in prep) indicate that the rates of these systems align with the inferred rate of lGRBs from observations.

Previous work in the modeling of these binary progenitor systems has provided many important details on whether or not the tidal interaction is significant enough to spin-up a star under a wide range of parameter space. \cite{izzard2004} examined the evolution and rates of rotating helium stars with and without a companion using the equations of \cite{Hurley2000, Hur02} to parameterize binary evolution. With this semi-analytic prescription, they find the companion is capable of spinning up the helium star to values of high angular momentum, and that the rates of these systems are high enough to be consistent with GRB progenitors. In a thorough investigation of mass and angular momentum loss due to a compact object companion, \citet{Det08} find that the best candidates for GRBs are Wolf-Rayet stars at low stellar metallicity, where stellar winds are weak and the orbital separations are stable without merger. While their study focuses on compact binaries that have undergone phases of mass transfer or common-envelope evolution, with orbital periods below $\sim$2 days, our parameter space is distinct: we consider significantly wider systems with initial orbital periods up to $\sim 10^5$ days and investigate how tidal interactions affect the stellar evolution and angular momentum of the MS.  Furthermore, \citet{Det08} consider stars at solar metallicity and we investigate MS under a broader range of metallicities ($10^{-4}-10^{-2}$).

\cite{Marchant2016} and \citet{Mandel2016} consider the interaction of MS in tight binaries under chemically homogenous evolution merger scenarios characterized by mixing and wind-driven mass loss due to rotation and tides. Additionally, there have been detailed studies looking at the influence of a low mass star on a MS in a tight binary, exploring the conditions under which the low mass star can spin-up the resultant BH that the MS produces upon collapse, but without merging \citep{MM22}. \citet{Qin18} investigate the angular momentum evolution of binary systems composed of a helium star and a BH companion with orbital periods shorter than two days. Their primary aim is to predict the spin of the second-born BH in systems that may eventually merge and be detectable via gravitational waves. As a secondary application, they briefly evaluate whether such systems could produce lGRBs. They conclude that lGRB production is only feasible in systems with very short orbital periods ($\leq$ 0.3 days), low metallicity, and efficient tidal synchronization. { In this work, we investigate a new region of binary parameter space by varying the component masses, stellar metallicities, and initial orbital periods up to $10^5$ days. We explore the limiting role of tidal interactions both with and without significant RLOF mass loss. We identify a subset of this space in which the MS retains sufficiently large spin angular momentum at the end of its lifetime that its angular momentum reservoir could plausibly power a GRB jet.}

%{  In this work, we investigate a new region of parameter space in our choice of binary masses, stellar metallicities, and initial orbital periods up to $10^5$ days while exploring the limit of tidal interactions with and without significant mass loss due to RLOF.  We highlight a range of parameter space in which the final spin angular momentum of the MS at the end of stellar lifetime is high enough, such that its angular momentum reservoir may be enough to power a GRB jet.}

   Our paper is organized as follows: In \S 2, we describe the code and simulation set-up (with a convergence study discussed in our Appendix).  In \S 3, we present our results and show the range of initial conditions that lead to a substantial spin-up of the MS in our systems. Additionally, we provide representative density profiles of the MS and briefly discuss the implications for the density profile of the circumbinary medium.  In \S 4, we present discussion and caveats to our results. In \S 5, we summarize our main conclusions and discuss future work to support this effort.   We use CGS units with gravitational constant $G$ and speed of light $c$.

\section{Code and Simulation Set-up}
\label{sec:code}
We use the one-dimensional stellar evolution code \emph{Modules for Experiments in Stellar Astrophysics} (\textsc{MESA}) for our simulations \citep{Paxton2011, Paxton2013, Paxton2015, Paxton2018, Paxton2019, paxton_bill_2021_5798242,Jermyn2023}\footnote{We use \textsc{MESA} release r21.12.1.}.  \textsc{MESA} solves the coupled structure and composition equations simultaneously.  The \textsc{MESA} equation of state (EOS) is a blend of the OPAL \citep{Rogers2002}, SCVH
\citep{Saumon1995}, FreeEOS \citep{Irwin2004}, HELM \citep{Timmes2000},
PC \citep{Potekhin2010}, and Skye \citep{Jermyn2021} EOSes.  Radiative opacities are primarily from OPAL \citep{Iglesias1993,
Iglesias1996}, with low-temperature data from \citet{Ferguson2005}
and the high-temperature, and the Compton-scattering dominated regime by
\citet{Poutanen2017}.  Electron conduction opacities are from
\citet{Cassisi2007} and \citet{Blouin2020}.  Nuclear reaction rates are from JINA REACLIB \citep{Cyburt2010}, NACRE \citep{Angulo1999} and
additional tabulated weak reaction rates \citet{Fuller1985, Oda1994,
Langanke2000}.  Screening is included via the prescription of \citet{Chugunov2007}.
Thermal neutrino loss rates are from \citet{Itoh1996}.  RL radii in binary systems are computed using the fit of
\citet{Eggleton1983}.  Mass transfer rates in RL
overflowing binary systems are determined following the
prescription of \citet{Ritter1988}.  { For rotation, the stellar structure equations in one dimension are solved under the ``shellular approximation,'' where the angular velocity is constant over isobars \citep{Paxton2013}.}

We employ the binary module in \textsc{MESA} because it enables simultaneous  {stellar }evolution of { a star and it's companion}, undergoing transfer of mass and angular momentum \citep{Paxton2015}.  We use the stellar rotation and tidal interaction treatment for the binary evolution in \citet{marchant2019} and \citet{selma_de_mink_2019}.  
With these tools, we investigate the regime where a BH companion significantly influences the evolution of a MS prior to collapse.  We describe the details of our simulation set-up below.

%%%%%%%%%%%%%%%%%%%%%%%%%%%%%%%%%%%%%%%%%%%%%%%%%%%%%%%%
\begin{table}%[h]
\centering
\caption{
 Initial conditions for MS-BH binary \textsc{MESA} simulations of circular orbits with initial BH mass $M_{\rm BH}$ in solar mass units $M_\sun$, initial MS mass $M_{*,i}$ in solar mass units $M_\sun$, stellar metallicity $Z$, initial MS radius $R_{*,i}$ in solar radius $R_\odot$ units, and initial orbital period $\tau_{\rm orb,i }$ in days.  
}
    \label{tab:simulation-parameters}
\begin{tabular}{| c | c | } 
\hline
Simulation Parameters & \\
\hline
\hline
  $M_{BH}[M_{\odot}]$ & \{10, 15\}\\
 \hline
 $M_{*, i}[M_{\odot}]$ & \{15, 25\}\\
  \hline
 $Z$ & \{1-9\}$\times 10^{-4}$, \{1-9\}$\times 10^{-3}, 10^{-2}$\\
 \hline
 $R_{*, i} [R_\odot]$ & \{4.4, $\cdots$, 6.0\}\\
 \hline
 $\tau_{\rm orb, i} [days]$ & 3 \ - \ 10$^5$\\
 %\hline
 %$a_{\rm orb, i}[R_\odot]$ & \\    
\hline
\end{tabular}
\end{table}

\subsection{MESA Input Physics}
\label{sec:flags} % used for referring to this section from elsewhere

 We evolve a Zero Aged Main Sequence (ZAMS) star in a circular orbit with a point mass BH at different mass ratios, orbital periods\footnote{We consider orbital periods of the BH that lead to significant tidal interaction but do not consider common envelope or merger scenarios.}, metallicities, stellar rotations, and mass/angular momentum transfer schemes.  %, and internal dynamo prescriptions.  

 Tab.~\ref{tab:simulation-parameters} gives { the} massive star -- black hole (MS-BH) binary simulation parameters for initial BH mass $M_{\rm BH}$, initial MS mass $M_{*,i}$, stellar metallicity $Z$, initial MS radius $R_{*,i}$, and initial orbital period $\tau_{\rm orb,i }$.  We present the most salient results from a suite of $\sim 2.3 \times 10^3$ \textsc{MESA} simulations.
 %, with the former indicating runs without initial rotation of the MS (i.e. all of the star's angular momentum is due to tidal interactions with the companion), and the latter shows our runs with an initial rotation of the MS along with a rotating convection flag turned on. 
 A short description of our initial set-up is as follows:\\
 
 \begin{itemize}
 \item Mass ratio -  We consider physically motivated binary systems, created in situ or by dynamical capture, to include an MS with an initial mass of $M_{*,i} [M_\odot] = \{15, 25\}$ \citep[a typical mass range used to model GRB progenitor stars, e.g.][]{WH06} and a BH companion of initial masses within the range of $M_{BH} [M_\odot] = \{10, 15\}$.  
 %Can refer to papers by Aleksey Generozov for dynamical capture.
 %We assume that the binary system is created in situ and not by dynamical capture.   Furthermore, we assume the BH companion is created from a star more massive that 25 $M_{\odot}$ such that it collapses first and does not lose so much mass upon collapse that it disrupts the binary system.
% we need that the BH is created from a MS that was more massive than 25 $M_{\odot}$ (so that it collapses first) but also one in which the collapse of this star did not lose so much mass upon collapse that it disrupts the binary system.
  As mentioned above, such systems, with these particular, { comparable} mass ratios, have been observed in X-ray binaries \citep{Kelley83, Lev93, Lev00, Tv06, Fal15}, and their rates are likely sufficient to account for the long GRB population \citep[see, e.g.][who examine these types of systems in the  context of their end-states as BH binaries, as well as A. Cason et al., in prep]{Bav21}.  In the latter study, we assume that the binary system is created in situ and not by dynamical capture, although including systems formed through dynamical capture will only bolster the rates.  For the purposes of this study, we do not focus on formation scenarios and/or population synthesis and assume these systems exist in sufficient numbers and focus on their individual evolution.
 
 \item Orbital Period -  For the MS-BH binary system, we define the orbital period $\tau_{\rm orb}$ as
 \begin{equation}
 \label{eq:orbital_period}
\tau_{\rm orb} = \sqrt{\frac{4\pi^2 a^3}{G M_{\rm tot}}},
 \end{equation}
 for a total mass, including the MS and BH, as $M_{\rm tot} = M_{*} + M_{\rm BH}$ and orbital separation $a$ \citep{Fitzpatrick2012}.
 In our study, we investigate a range of orbital periods (and subsequently orbital separations) where the BH companion may tidally influence the MS.  { At orbital separations larger than the stellar radius, a Newtonian gravity treatment to the binary interaction is sufficient because general relativistic corrections to the orbit and tide are very small.}  We are primarily interested in the influence of the companion at a distance outside of the Roche lobe (RL) radius of the MS.  We define the RL radius $R_{\rm RL}$ for the MS as,
\begin{equation}
\label{eq:rochelobe}
    R_{\rm RL} = \left [ \frac{0.49 q_1^{2/3}}{0.6 q_1^{2/3} + \ln(1+q_1^{1/3})}\right ] a, 
\end{equation} 
for binary mass ratio $q_1 = M_*/M_{\rm BH}$ \citep{Paxton2015}.  Mass transfer or RL overflow (RLOF) occurs when the stellar radius $R_*$ approaches or exceeds the RL radius or $R_*/R_{\rm RL} \gtrsim 1$.  For the simulation parameters given in Tab.~\ref{tab:simulation-parameters}, the closest binary encounter or smallest initial orbital period is three days. { This corresponds to distances outside of the Roche lobe radius, with initial orbital separations of $\sim 25\text{–}30,R_\odot$ and MS Roche lobe radii of $\sim 10\text{–}13,R_\odot$ across the binary mass combinations considered.}

%{ This corresponds to distances outside of of the RL radius with initial orbital separations of $\sim 25-30 R_\odot$ and MS RL radii at $\sim 10-13 R_\odot$ for the given combinations of binary masses.}  

 \item Metallicity -  We consider a range of values between $Z=10^{-4}$ and $Z=10^{-2}$, where metallicity ($Z$) is defined as the mass fraction elements heavier than helium relative to the total mass of the gas.  We note that these span two orders of magnitude but remain lower than solar metallicity ($\sim$$1.7\times 10^{-2}$), 
  %0.017$)
 { since} it has been demonstrated that long GRBs tend to occur on average in lower metallicity environments \citep{Fru99,LF03, F06, Lev10, GF13, GF17, Pal19, GSF19}. These values provide the endpoints to a large dynamic range over which to explore dependencies on metallicity.   The corresponding range in initial stellar radius is $R_{*, i} \approx 4-6 R_\odot$.
 
\item Spin Synchronization - The binary module in MESA assumes that the star’s rotational axis is perpendicular to the orbital plane, enabling tidal interactions to contribute to the star’s spin-up through angular momentum transfer.  MESA uses a modified scheme to solve for the star’s angular frequency as a function of time \citep{Hut1981, Hur02, Paxton2015}, by solving for the evolution in close binaries which includes tidal interactions, mass transfer, stellar winds, and common-envelope evolution. The time evolution of the angular frequency of the MS is given by  
\begin{equation}
\frac{d\Omega_{i}}{dt} = \frac{\Omega_{\rm orb} - \Omega_{i}}{\tau_{\rm sync}},
%   \frac{d\Omega_{i}}{dt} = \frac{3}{ \left( q_{}^2 r_{g}^2\right)} \left( \Omega_{\mathrm{orb}} - \Omega_{i} \right) \left(\frac{k}{T}\right)_{c} \left(\frac{R}{a} \right)^6
  %\left( \frac{q_1^2}{r_{g}^2} \right) \left( \frac{R}{a} \right)^6
\label{eq:spin_torque}
\end{equation}
where $\Omega_i$ is the angular frequency at the face of cell \(i\) toward the surface and $\tau_{\rm sync}$ is the tidal synchronization time-scale given by
\begin{equation}
\label{eq:tsync}
\tau_{\rm sync} = 52^{-5/3} \frac{t_* r_g^2}{E_2} q_2^{-2} (1+q_2)^{-5/6} \left (\tfrac{a}{R_*} \right )^{17/2},
\end{equation}
where $q_2 = M_{\rm BH}/M_*$, $t_* = \sqrt{R_*^3/(GM_*)}$, $r_{g}^2 = I_*/(M_*R_*^2)$ for stellar moment of inertia $I_*$, and $E_2 = 1.592\times 10^{-9} (M_*/M_\odot)^{2.84}$ for MS \citep{Zahn1975,Hut1981, Hur02, Paxton2015}.  Similar to \citet{Det08}, \textsc{MESA} evolves Eq.~\ref{eq:spin_torque} by assuming constant $\tau_{\rm sync}$ and $\Omega_{\rm orb}$ through a time step $\delta t$ where $\Delta\Omega_i = [1 - \exp(-\delta t/\tau_{\rm sync})](\Omega_{\rm orb} - \Omega_i)$ \citep{Paxton2015}.
%
%$r_{g}^2 = I/(MR^2)$   { is the radius of gyration and  \textit{I} is the moment of inertia of the MS. The ratio of the apsidal motion constant \( k \), which describes the internal structure of the star, to the damping time-scale \( T \), which characterizes how quickly tidal interactions alter the orbit, is denoted as \( k/T \). When calculating the ratio \( k/T \), MESA follows the formalism of \citet{Hur02}, who developed a rapid binary evolution algorithm that incorporates tidal interactions, mass transfer, stellar winds, and common-envelope evolution, including a detailed treatment of tidal evolution using \( k/T \) to quantify tidal effects.
%Lastly, the ratio of the star's radius to the orbital separation is denoted as \(R/a\).  
For all simulations, we do not initially relax the stellar rotation to the orbital period.

 \item MS initial rotation - We consider a range of stable, initial rotations for the MS under the { upper} limit of stellar break-up velocity.  We initialize MS rotation at a velocity given by the binary evolution equations in \cite{HPT00} and \cite{Hur02}, used in the binary population synthesis code COSMIC \cite{Brev20}.
 We consider zero velocity $(v_{0} )$ as a standard point in investigating the effects of spin.  We use two additional criteria for the initial velocity.  The first ($v_1$) = 9.177 $\times 10^1$ km/s ) is a representative initial condition from binary simulations in \cite{Kenoly23} using the \textsc{COSMIC} population synthesis code \citep{Brev20}. { The second is 
 half of the critical velocity $(0.5v_{*})$,
defined as an estimate of the velocity at which the star's rotation rate will cause break up, $v_* \sim \sqrt{GM_*/R_*}$.}

 \item Mass transfer -  We consider tidal interactions between the MS and BH where the BH companion is initially at a distance beyond the RL of the MS given in Eq.~\ref{eq:rochelobe}.  In \textsc{MESA}, we  employ a Kolb-Ritter \citep{KR90,Paxton2015} mass transfer scheme, which accounts for stellar parameters such as finite atmosphere scale height and detailed structure of the outer layers.  Mass transfer attributed to RLOF is given by
\begin{equation}
    \begin{split}
    \dot{M}_{\rm RLOF} = -\dot{M}_0 - 2\pi F_1(q_2) \frac{R_{\text{RL}}^3}{G M_*}
    \times \\
    \int_{P_{\text{ph}}}^{P_{\text{RL}}} 
    \Gamma_1^{1/2} \left( \frac{2}{\Gamma_1 + 1} \right)^{\frac{\Gamma_1 + 1} 
    {2(\Gamma_1 - 2)}}
    \left( \frac{k_B T}{m_p \mu} \right) ^{1/2} dP,
    \end{split}
    \label{eq:KR-scheme}
\end{equation}
 for pressure $P$, temperature $T$, and mean molecular weight $\mu$, where the mass transfer rate of the MS in the binary is
\begin{equation}
    \dot{M}_0 = \frac{2 \pi}{\exp(1/2)} F_1(q_2) \frac{R_{\text{RL}}^3}{G M_*} \left( \frac{k_B T_{\text{eff}}}{m_p \mu_{\text{ph}}} \right)^{1/2} \rho_{\text{ph}},
\end{equation}
 and fitting function is
\begin{equation}
    F_1(q_2)=1.23 + 0.5 \log q_2 \hspace{2mm}, \hspace{2mm} 0.5 \lesssim q_2 \lesssim 10,
\end{equation}
 for first adiabatic exponent $\Gamma_1$, density at the photosphere $\rho_{\rm ph}$, pressure at the photosphere $P_{\rm ph}$,  pressure at the RL radius $P_{\rm RL}$, Boltzmann constant $k_B$, effective temperature $T_{\mathrm{eff}}$, proton mass $m_p$, and mean molecular weight at the photosphere $\mu_{\rm ph}$.   Furthermore, we limit the mass transfer to be lower than the Eddington limit.

 There are many subtleties when choosing the mass transfer prescriptions.  For example,  \cite{CP23} find an intermediate solution between optically thin and optically thick regimes for RLOF; their model finds a mass transfer rate about two times smaller than \cite{KR90} and four times smaller than \cite{Marchant2016} for a 30 $M_{\sun}$, low metallicity star.  
 %However, again, because we are considering encounters where the initial separation distances are larger than the Roche lobe radius, these subtleties generally do not influence our main results.
 We note the recent study by \cite{Klen25} who show that there is a lower limit to the orbital separation in MS binary systems, in order to ensure stable mass transfer (and prevent a common envelope phase, which can be extremely unstable and in which the two stars will likely rapidly merge).  They find this orbital separation to be $\gtrsim 10 R_{\odot}$.

  In our simulations non-conservative mass transfer occurs when mass from the MS due to winds or RLOF can remove angular momentum from the system.  In \textsc{MESA}, the corresponding loss in angular momentum is calculated from the net mass loss \citep{Paxton2015,marchant2019,selma_de_mink_2019}.

 \item Spruit-Tayler Dynamo - 
 %We run simulations with and without a Spruit-Taylor Dyanmo \cite{Spruit2002} flag.  
Through magnetic coupling between layers in the stellar interior, the Spruit-Tayler dynamo \citep{Spruit2002,Fuller2019,Ma2019} is a possible pathway to transport angular momentum within the MS, potentially leading to a spin down.  However, as we discuss below, its effect is more pronounced in stars less massive than those we consider here  \citep[see also the analytic arguments and discussion in][who show that for slowly spinning MS the slow-down from a Spruit-Taylor dynamo may be avoided]{MM14}. Nonetheless, we have run simulations with this flag both on and off and find a negligible impact on our results (we discuss this further below).

 \item Rotation in Convection Zone - %We use MESA's default flag for rotation in the convection zone to further test what could influence the MS's angular momentum. 
% In the case of rotating convection, there are many complexities that arise in these models. 
In the case of rotation in the convection zone, angular momentum can be be transported through various processes internal to the star \citep{Augustson_2019}. These processes are controlled by the diffusion coefficients of Goldreich-Schubert-Fricke (GSF), Eddington-Sweet (ES) circulation, secular shear instability (SSI), and dynamical shear instability(DSI). The GSF \citep{gold_schu,fricke1968instabilitat} coefficient quantifies how angular momentum is transported in the interiors of stars. ES occurs when there are imbalances of pressure and temperature inside the interior of the star.  SSI is a long-term instability in rotating stars where thermal diffusion weakens stratification, allowing shear-driven turbulence to redistribute angular momentum and mix elements.  DSI occurs in rotating stars when shear forces exceed stabilizing buoyancy, triggering turbulence and rapid mixing on a dynamical timescale \citep{Heger_2000}. 

 \item Termination condition - We identify the end point for our simulations as the stage prior to collapse, when the carbon-oxygen core forms at $\sim$6-10 Myr with a corresponding rapid increase in central temperature and density within the star.   The final time delineated by the formation of the CO core is consistent with the GRB progenitor scenario discussed in \citet{izzard2004}. {   %We track the evolution of the helium core, carbon-oxygen core, and central mass fraction of helium-4 and note that our stopping criteria for the simulations is consistent with this end point.  
 In the Appendix Sec.~\ref{sec:convergence}, we characterize stellar termination as the coincident time where the helium and carbon-oxygen cores form and the central mass fraction of helium-4 decreases to $\lesssim 10^{-3}$.  With these conditions, we present the final value of the spin angular momentum as well as the total mass loss in our simulations in  Fig.~\ref{fig:main-results2} and Fig.~\ref{fig:main-results3} for binary parameter space listed in Tab.~\ref{tab:simulation-parameters}.}

% We report the final angular momentum transfer to the star at this stage in the simulated evolution. 

 \item Wind Mass Loss - We adopt the isotropic `Dutch' wind mass loss scheme in MESA, where MESA automatically applies phase-dependent mass loss prescriptions throughout the evolution of the star. MESA follows the formulation of \citet{glebbeek}, where they use the theoretical mass values for stars with effective temperatures between 10,000 K $\leq$ $T_{eff}$ $\leq$ 50,000 K from \citet{vink2000,vink2001}. Then, when $T_{eff}$ falls below the lower bound, the empirical method from \citet{jager1988} is implemented. Both of these methods account for various stages in stellar evolution.
 \item Evolution of Orbital Separation -  Mechanisms related to mass loss alter the binary's orbital separation both by removing mass and angular momentum.  One may use the relation in \citet{tauris_2006} to determine the evolution of the orbital separation of the binary and binary parameters as 
\begin{equation}
    \begin{split}
    \frac{\dot{a}}{a} = 2 \frac{\dot{J}_{orb}}{J_{orb}} - 2 \frac{\dot{M}_*}{M_*}  - 2 \frac{\dot{M}_{\rm BH}}{M_{\rm BH}} + \frac{\dot{M}_* + \dot{M}_{\rm BH}}{M_{\rm tot}}, 
%     \\ M = M_1 + M_2
    \end{split}
    \label{eq:orbital_angmom-balance}
\end{equation}
which depends on changes in the system's orbital angular momentum, $\dot{J}_{\rm orb}$, and the mass loss rates of the MS and BH, $\dot{M}_*$ and $\dot{M}_{\rm BH}$, respectively.
%, with $M_1$ and $M_2$ representing their respective masses.}
%\textcolor{red}{Cite the reference for this equation.}}
\end{itemize}

%%%%%%%%%%%%%%%%%%%%%%%%%%%%%%%%%%%%%%%%%%%%%%%%%%%%%%%
\begin{figure*}
    \centering
 \includegraphics[width = 0.495\textwidth]{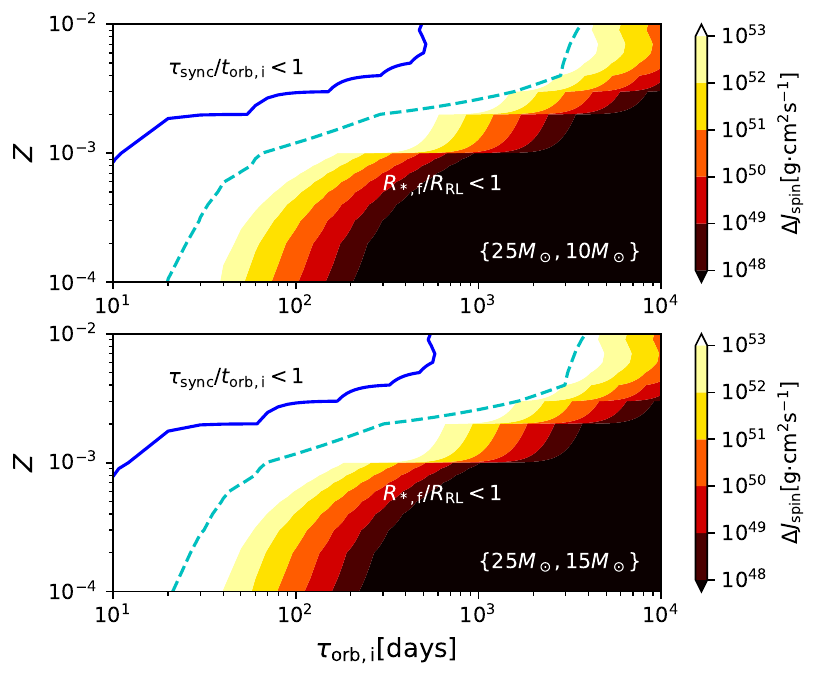}
  \includegraphics[width = 0.495\textwidth]{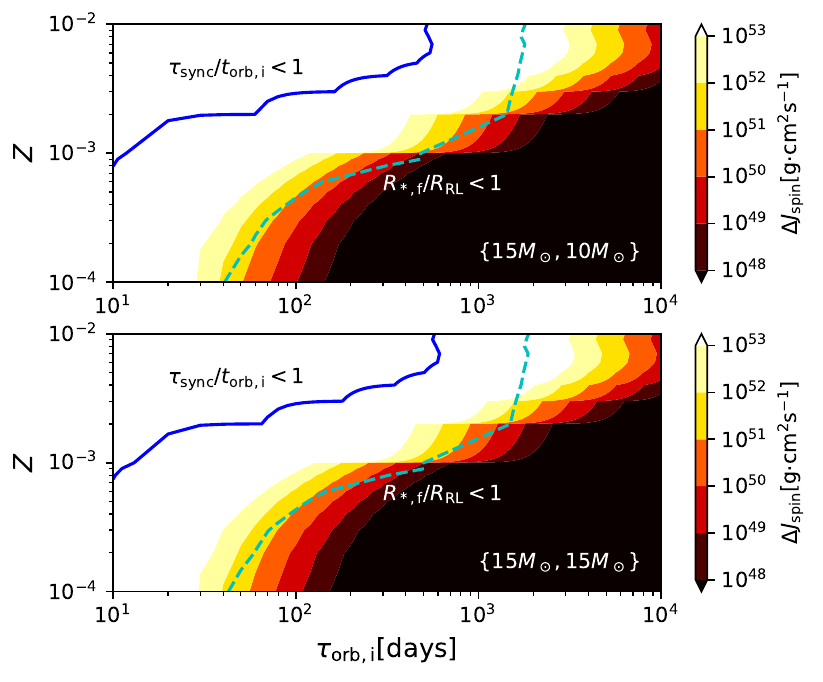}
 \caption{Estimate of the stellar spin-up in massive star -- black hole binary (MS-BH) systems under a range of initial orbital periods and metallicities for { an initially non-rotating MS.}  We give spin angular momentum or stellar spin-up $\Delta J_{\rm spin}$ in CGS units.  We consider the parameter space presented in Tab.~\ref{tab:simulation-parameters} for MS of initial masses $M_{*,i} [M_\odot] = \{15, 25\}$ under a range of metallicities of $Z=10^{-4}-10^{-2}$ and BH of masses $M_{BH} [M_\odot] = \{10, 15\}$.  The initial orbital period of the MS-BH binary, $t_{\rm orb, i}$, is given in days.  In the dashed cyan curve, we give the limit where the ratio of final stellar radius to RL overflow radius $R_{*,f}/R_{\rm RL}$ is unity.  { We expect mass loss due to RLOF in the region of parameter space approximately to the left of this curve, where $R_{*,f}/R_{\rm RL} > 1$.  To the right of this curve, we expect negligible mass loss, where $R_{*,f}/R_{\rm RL} < 1$.  In the solid blue curve, we give the limit where the ratio of tidal synchronization $\tau_{\rm sync}$ %given in Eq.~\ref{eq:tsync} 
 to initial orbital timescale  $\tau_{\rm sync}/t_{\rm orb, i}$ is unity.  We expect tidally synchronized binary orbits in the region of parameter space approximately to the left of this curve, where $\tau_{\rm sync}/t_{\rm orb, i} < 1$.   To the right of this curve, we expect that these binaries may not be tidally synchronized over a single orbit, where $\tau_{\rm sync}/t_{\rm orb, i} > 1$.}
 }
    \label{fig:main-estimate}
\end{figure*}
%%%%%%%%%%%%%%%%%%%%%%%%%%%%%%%%%%%%%%%%%%%%%%%%%%%%%%%

\subsection{Theoretical expectations}
\label{subsec:theoretical_expectations}
 In our investigation, we simulate the interplay of physical processes in stellar and binary evolution using the numerical tools described in Sec.~\ref{sec:flags}.  We argue that by angular momentum conservation, the main mechanisms that determine the { final spin} of the MS are those that provide significant changes to the orbital parameters of the binary and stellar mass loss.  While individual mechanisms such as stellar wind mass loss in single star evolution or tidal interactions may play a significant role, we consider the combined, non-linear effects of these mechanisms along with other forms of angular momentum loss such as magnetic coupling and rotation in the binary simulations.  In the following, we approximately identify the region of parameter space in initial orbital period $\tau_{\rm orb, i}$ of the binary and metallicity $Z$ of the MS where tidal interactions significantly increase the spin angular momentum of the star.  We estimate the time rate of change of the average stellar spin angular momentum $J_{\rm spin}$ by relating the tidal torque $\tau$ by the orbiting BH about the center of the MS to the average stellar angular frequency $\Omega_*$ and moment of inertia $I_*$ with
\begin{equation}
\label{eq:tidal_torque}
    \tau = I_* \frac{d\Omega_*}{dt} = \frac{dJ_{\rm spin}}{dt}.
\end{equation}
%

% =============

 { Then, we approximate the stellar spin-up $\Delta J_{\rm spin}$ due to the binary tidal interaction within a time duration $\Delta t$ for an initially non-rotating MS using the time evolution of the angular frequency (Eq.~\ref{eq:spin_torque}) and tidal synchronization time-scale (Eq.~\ref{eq:tsync}) in Eq.~\ref{eq:tidal_torque}, where
\begin{equation}
\label{eq:maximum_spin_estimate}
   \Delta J_{\rm spin} \approx I_*  \frac{\Omega_{\rm orb}}{\tau_{\rm sync}} \Delta t = 2 \pi \frac{I_*}{\tau_{\rm sync}}\frac{\Delta t}{\tau_{\rm orb}},
\end{equation}
for orbital frequency $\Omega_{\rm orb} = 2 \pi /\tau_{\rm orb}$ and stellar moment of inertia $I_* \approx (2/5) M_* R_*^2$. For a tidally locked binary system, the spin angular momentum of the MS prior to collapse can also be estimated analytically \citep{LR22} as
\begin{equation}
  \label{eq:spin-up-binary}
      J_{\rm spin} \sim M_* R_*^{2} \left(\frac{G M_{\rm tot}}{a_{\rm orb}^{3}} \right)^{1/2},
\end{equation} 
for stellar mass $M_*$, stellar radius $R_*$, orbital separation $a_{\rm orb}$, and total binary mass $M_{\rm tot} = M_{\rm BH} + M_*$. The spin-up of an initially non-rotating MS in our estimate is consistent with this analytic scaling in the regime where the binary is efficiently tidally synchronized.
Identifying $I_* \approx (2/5) M_* R_*^2$ and taking $\Delta t \sim \tau_{\rm sync}$, Eq.~\ref{eq:maximum_spin_estimate} then gives $\Delta J_{\rm spin} \sim I_* \Omega_{\rm orb}$, which has the same dependence on $M_*$, $R_*$, and $a_{\rm orb}$ as Eq.~\ref{eq:spin-up-binary}. In practice, we use Eq.~\ref{eq:maximum_spin_estimate} as a torque-based estimate for $\Delta J_{\rm spin}$, with Eq.~\ref{eq:spin-up-binary} providing a useful synchronized limiting case.}

%Then, we approximate the stellar spin-up $\Delta J_{\rm spin}$ due to the binary tidal interaction within a time duration $\Delta t$ for an initially non-rotating MS using the time evolution of the angular frequency (Eq.~\ref{eq:spin_torque}) and tidal synchronization time-scale (Eq.~\ref{eq:tsync}) in Eq.~\ref{eq:tidal_torque}, where
%
%\begin{equation}
%\label{eq:maximum_spin_estimate}
%   \Delta J_{\rm spin} \approx I_*  \frac{\Omega_{\rm orb}}{\tau_{\rm sync}} \Delta t = 2 \pi \frac{I_*}{\tau_{\rm sync}}\frac{\Delta t}{\tau_{\rm orb}},
%\end{equation}
%for orbital frequency $\Omega_{\rm orb} = 2 \pi /\tau_{\rm orb}$ and stellar moment of inertia $I_* \approx (2/5) M_* R_*^2$.  {For a tidally locked binary system, the spin angular momentum of the MS prior to collapse can also be estimated analytically \citep{LR22} as
%
%\begin{equation}
%  \label{eq:spin-up-binary}
%      J_{\rm spin} \sim M_* R_*^{2} \left(\frac{G M_{\rm tot}}{a_{\rm orb}^{3}} \right)^{1/2},
%\end{equation} 
%
%for stellar mass $M_*$, stellar radius $R_*$, orbital separation $a_{\rm orb}$, and total binary mass $M_{\rm tot} = M_{\rm BH} + M_*$. The spin-up of the MS in our simulations is consistent with this analytic scaling in the regime where the binary is efficiently tidally synchronized. }

%==============

We choose $\Delta t$ as the time duration of significant torque due to the stellar radius appreciably increasing at the end of the MS lifetime.  We consider a value of $\Delta t \sim 10^{4}$ yr  as a maximum estimate in surveying simulation results of single star evolution at a range of metallicities.  This increase in MS stellar radius towards the end of its lifetime is discussed in Appendix Sec.~\ref{sec:appendix_singlestar}.  Furthermore, we note that the stellar radius or metallicity is constrained by the stellar evolution model in \textsc{MESA} or $R_*(M_*,Z)$.  We then obtain a scaling dependence for the stellar spin-up $\Delta J_{\rm spin}$ on initial orbital period and stellar metallicity or $\Delta J_{\rm spin} = \Delta J_{\rm spin} (M_{\rm BH}, M_*, Z, t_{\rm orb, i})$ using the tidal synchronization time-scale in terms of the mass ratio and orbital period of the binary.  
%%%%%%%%%%%%%%%%%%%%%%%%%%%%%%%%%%%%%%%%%%%%%%%%%%%%%%%
\begin{figure*}
    \centering
 \includegraphics[width = 0.495\textwidth]{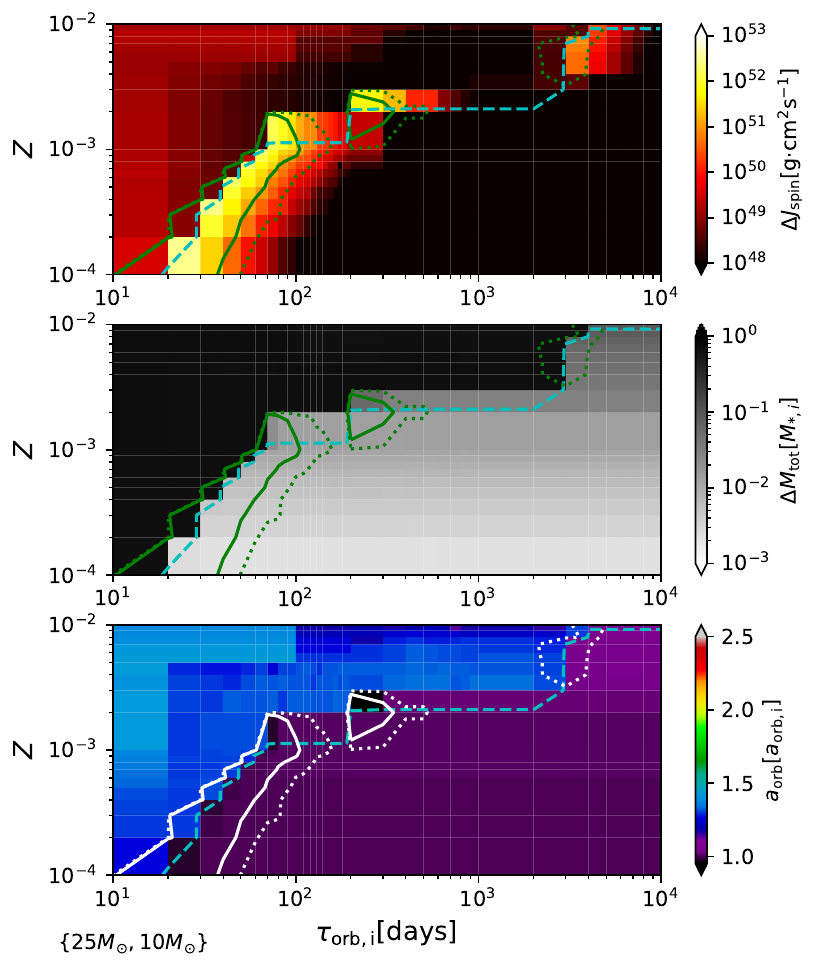}
  \includegraphics[width = 0.495\textwidth]{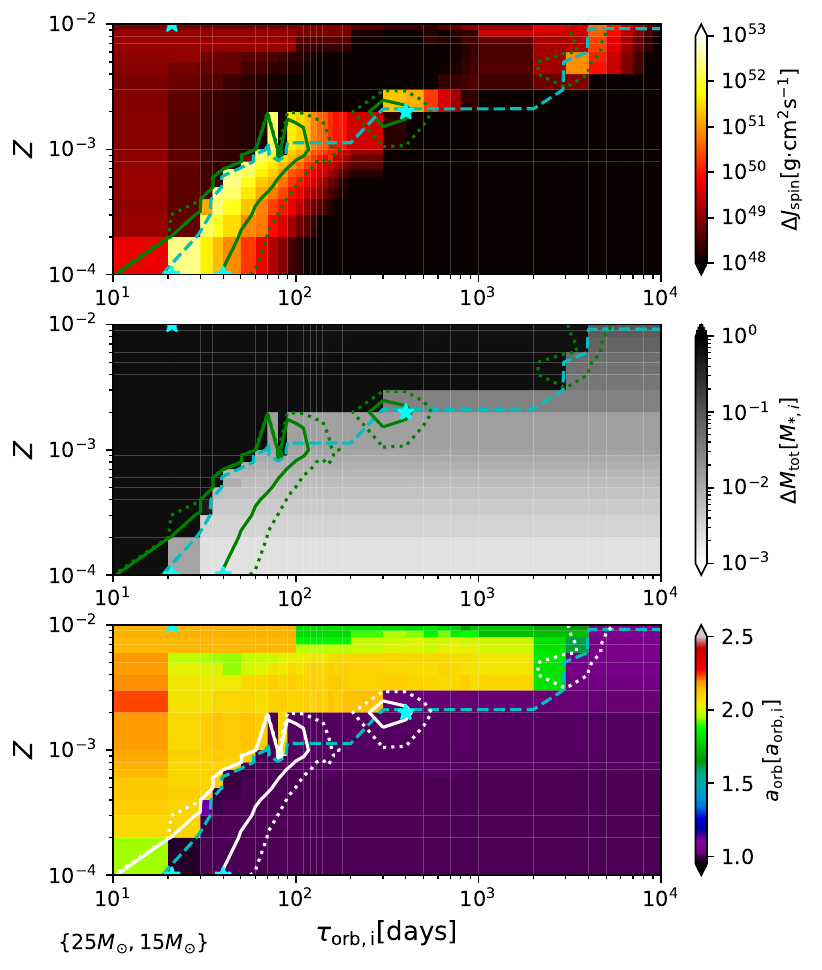}
 \caption{ \textsc{MESA} simulation results of the final stellar spin-up in MS-BH systems under a range of initial orbital periods and metallicities for { an initially non-rotating} MS.  We compare parameter studies of a MS with initial mass $M_{*,i}$ = 25$M_\odot$, while varying the BH mass $M_{BH} [M_{\odot}] = \{10, 15\}$ in left and right columns, respectively.   Initial orbital period $t_{\rm orb, i}$ is given in units of days.  In the top row, we give the final change in spin angular momentum or stellar spin-up $\Delta J_{\rm spin}$ in CGS units at the end of stellar lifetime.  In the middle row, we give the corresponding total mass loss $\Delta M_{\rm tot}$ in units of initial mass $M_{*,i}$. In the bottom row, we give the orbital separation $a_{\rm orb}$ in units of initial orbital separation $a_{orb,i}$.  {  We identify the region of parameter space of possible GRB progenitors as inside the solid (dotted) contour line that is green for the top and middle panels and white for the bottom panel.  This line approximately represents the value $10^{51} (10^{50}) {\rm g}\cdot{\rm cm}^2 {\rm s}^{-1}$, the estimate for the angular momentum of a post-collapse BH mass of $10 M_\odot$ ($5 M_\odot$) and spin of 1.0 (0.5).  In the dashed cyan line, for all three panels, we show the approximate contour that indicates a mass loss of 10\% of the initial mass of the star as the boundary of significant mass loss. For reference, in the right column, we show cyan star markers which correspond to specific simulation parameters used in a focused study of evolution in stellar and binary orbital parameters described in Sec.~\ref{sec:results} and Fig.~\ref{fig:evol21day}-\ref{fig:initial_rot2}.  }
 }    \label{fig:main-results2}
\end{figure*}
%%%%%%%%%%%%%%%%%%%%%%%%%%%%%%%%%%%%%%%%%%%%%%%%%%%%%%%
\begin{figure*}
    \centering
 \includegraphics[width = 0.495\textwidth]{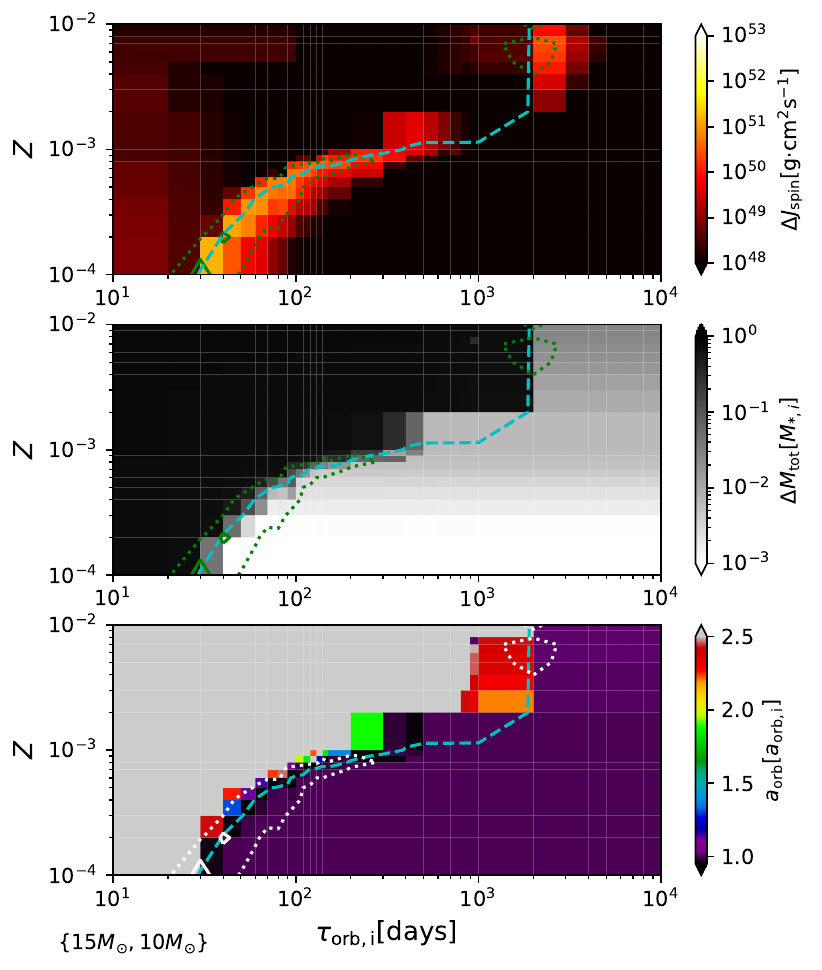}
  \includegraphics[width = 0.495\textwidth]{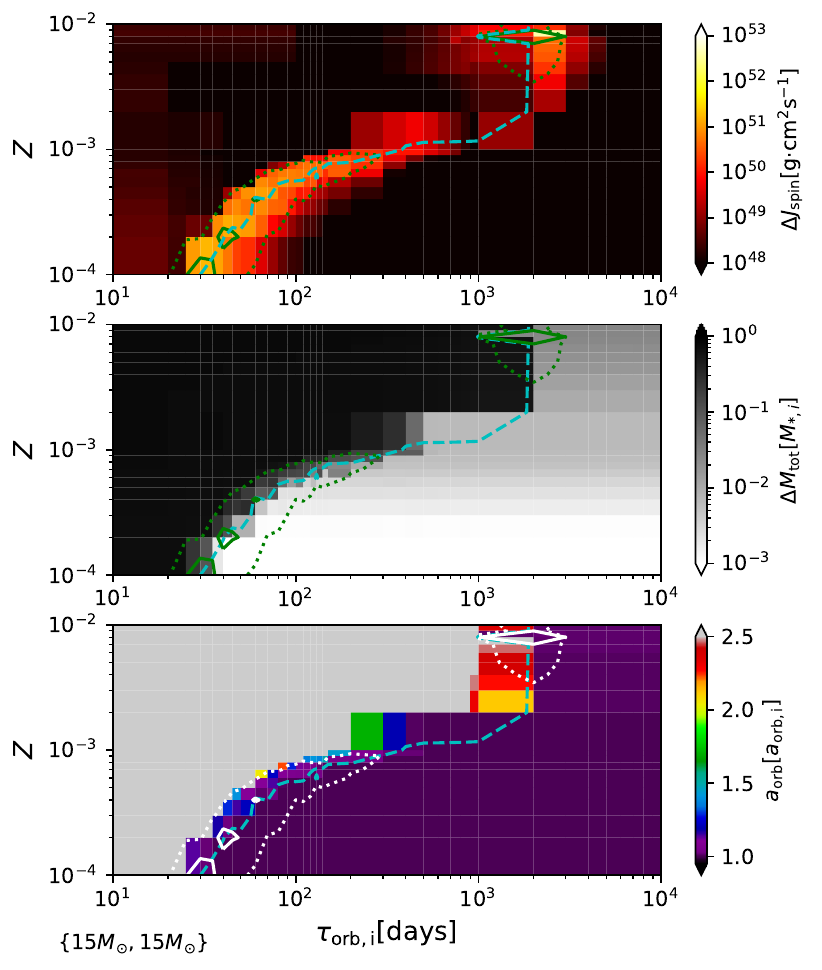}
 \caption{\textsc{MESA} simulation results of the final stellar spin-up in MS-BH systems under a range of initial orbital periods and metallicities for { an initially non-rotating} MS.  We compare parameter studies of a MS with initial mass $M_{*,i}$ = 15$M_\odot$, while varying the BH mass $M_{BH} [M_{\odot}] = \{10, 15\}$ in left and right columns, respectively.   Initial orbital period $t_{\rm orb, i}$ is given in units of days.  In the top row, we give the final change in spin angular momentum or stellar spin-up $\Delta J_{\rm spin}$ in CGS units at the end of stellar lifetime.  In the middle row, we give the corresponding total mass loss $\Delta M_{\rm tot}$ in units of initial mass $M_{*,i}$. In the bottom row, we give the orbital separation $a_{\rm orb}$ in units of initial orbital separation $a_{orb,i}$. {  We identify the region of parameter space of possible GRB progenitors as inside the solid (dotted) contour line that is green for the top and middle panels and white for the bottom panel.  This line approximately represents the value $10^{51} (10^{50}) {\rm g}\cdot{\rm cm}^2 {\rm s}^{-1}$, the estimate for the angular momentum of a post-collapse BH mass of $10 M_\odot$ ($5 M_\odot$) and spin of 1.0 (0.5).  In the dashed cyan line, for all three panels, we show the approximate contour that indicates a mass loss of 10\% of the initial mass of the star as the boundary of significant mass loss. }}
    \label{fig:main-results3}
\end{figure*}
%%%%%%%%%%%%%%%%%%%%%%%%%%%%%%%%%%%%%%%%%

In Fig.~\ref{fig:main-estimate}, we estimate the stellar spin-up for the MS with the scaling dependence for $\Delta J_{\rm spin}$.  We use the final mass $M_{*, f}$ and radius $R_{*, f}$ of a single star evolved until collapse with \textsc{MESA} under the termination conditions described in Sec.~\ref{sec:flags} and Appendix Sec.~\ref{sec:appendix_singlestar}.  Fig.~\ref{fig:main-estimate} gives the spin $\Delta J_{\rm spin}$ in CGS units under a range of initial orbital periods $\tau_{\rm orb, i}$ and metallicities $Z$ for the simulation parameters given in Tab.~\ref{tab:simulation-parameters}. { We consider binary interactions of a MS-BH system in which the initial mass of the MS and BH are $M_{*,i} [M_\odot] = \{15, 25\}$ and $M_{\rm BH} [M_\odot] = \{10, 15\}$, respectively and correspond to comparable mass ratios, $M_*/M_{\rm BH} = \{1.0, 1.5, 1.667, 2.5\}$.}  Using Eq.~\ref{eq:rochelobe} for $R_{\rm RL}$ at final stellar mass $M_{*, f}$, we give the limit where the ratio of final stellar radius to RL overflow radius $R_{*,f}/R_{\rm RL}$ is unity in the dashed cyan curve.  Approximately to the left of this curve, we expect mass loss due to RL overflow in the region of parameter space.  To the right of this curve, the ratio is less than unity or $R_{*,f}/R_{\rm RL} < 1$.  We note that in our estimate of $\Delta J_{\rm spin}$, we neglect angular momentum loss due to RLOF although significant mass loss ($\gtrsim 0.1 M_{*,i}$) will impact the final stellar spin angular momentum.  Therefore, we consider reliable estimates of $\Delta J_{\rm spin}$ to be in the parameter space in the limit of $R_{*,f}/R_{\rm RL} < 1$ (right of the cyan dashed curve).  
{ In the solid blue curve, we show the limit where the ratio of tidal synchronization timescale, given in Eq.~\ref{eq:tsync}, to initial orbital timescale $\tau_{\rm sync}/t_{\rm orb, i}$ is unity.   We expect tidally synchronized binary orbits in the region of parameter space approximately to the left of this curve, where $\tau_{\rm sync}/t_{\rm orb, i} < 1$.   To the right of this curve, we expect that these binaries may not be tidally synchronized over a single orbit, where $\tau_{\rm sync}/t_{\rm orb, i} > 1$. We note that the binary interaction is over numerous orbits over the lifetime of MS and this may lead to synchronization.  For a stellar lifetime of $\sim 6-10$ Myr, the number of binary orbits corresponding to the parameters in Tab.~\ref{tab:simulation-parameters} range from $\sim 10^4-10^9$.  Furthermore, as we have discussed, as the stellar radius increases at the end of the MS lifetime, the binary orbits at this stage will contribute a significant amount of torque.  An estimate of how much the tidal interaction will synchronize the orbit is outside the scope of the order of magnitude estimate here and best captured by the numerical investigation below.  We do anticipate that shorter initial orbital periods will likely synchronize orbits more than longer initial orbital periods due to the respective number of binary orbits completed before the end of stellar lifetime. 
}
%For the estimate in Fig.~\ref{fig:main-estimate}, we choose $t_f$[yr] = \{$1.3 \times 10^7, 7.5 \times 10^6$\}  for $M_{*,i} [M_\odot] = \{15, 25\}$ respectively.  

{ The estimates in Fig.~\ref{fig:main-estimate} across all binary pairs of MS and BH masses are morphologically similar in the parameter space.  }For a given MS initial mass, in comparing results between BH masses (top row versus bottom row), the differences are small.  { We find a greater dependence on the choice of masses in the location of the RLOF limit, where $R_{*,f}/R_{\rm RL}$ is unity, between the lower and higher ranges of mass ratio (left versus right column) due to the initial mass of the MS.  Furthermore, within the region of inappreciable mass loss ($R_{*,f}/R_{\rm RL} < 1$), the magnitude of final stellar spin-up $\Delta J_{\rm spin}$ is greater and region of relevant parameter space is wider for simulations with a MS initial mass of $25M_\odot$ than for a MS initial mass of $15M_\odot$.}  We expect the main mechanisms contributing to the final spin-up are the tidal interaction and stellar wind if the $\textsc{MESA}$ simulation results share morphological features in the parameter space of Fig.~\ref{fig:main-estimate}.

\subsection{Identification of viable binary systems}
In the following, we present our set of binary parameters that could lead to potential progenitors of radio-bright lGRBs, particularly through significant stellar spin-up over the MS lifetime due to interaction with its companion.   Our definition of "viable" means that the star has enough angular momentum at the end of its lifetime to power a GRB jet (assuming that at least ten percent of the angular momentum at the end of the star's life is retained in the BH disk system, allowing it to launch a sufficiently powerful jet to create a GRB; we discuss the details and caveats of this assumption further in our Discussion below).  In our numerical simulations, we track the evolution of quantities that describe the dynamics of the binary system such as stellar and BH mass, orbital separation of the binary, and change in spin, orbital, and total angular momentum as well as stellar quantities  such as mass loss due to wind and RLOF, stellar radius, RL radius, and rotation. Low-metallicity environments are further singled out as observationally favored for collapsar production, since smaller radii and reduced wind losses yield more massive, rapidly rotating cores \citep{izzard2004}. We report the spin angular momentum, total mass loss, and orbital separation for the MS at the final stage of evolution indicated by our stopping criteria.  
%through radial profiles in density and temperature.  
In the Appendix Sec.~\ref{sec:convergence}, we discuss convergence studies of representative models of the binary interaction between a $15 M_\odot$ BH and $25 M_\odot$ MS at an initial orbital period of three days to test our schemes where significant mass loss occurs.  We show that our quantities of interest do not change appreciably over stellar lifetime and model number at increasing mass resolution.

\section{Results} 
\label{sec:results}

\begin{figure*}
    \centering
\includegraphics[width = 0.95\textwidth]{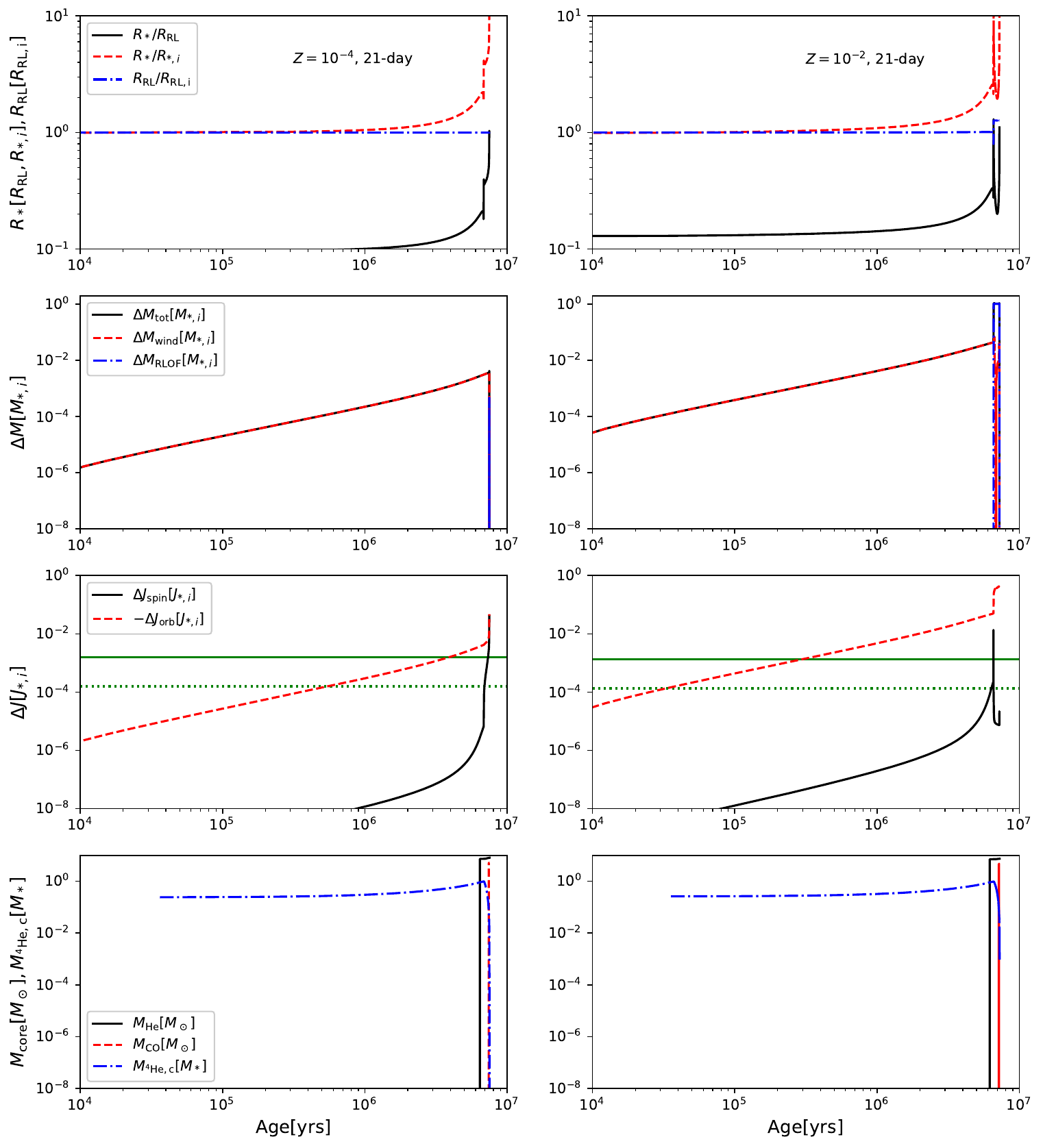}
\caption{  { Focused study of the impact of stellar metallicity on mass loss and final stellar spin-up of an initially non-rotating MS at stellar termination.  We show the evolution of a MS interacting with a 15$M_\odot$ BH at an initial orbital period of twenty-one days.  }  Stellar age is given in units of years.  We compare simulation results of a MS of initial mass $M_{*,i}$ = 25$M_\odot$ with metallicity $Z=10^{-4}$ and $Z=10^{-2}$ in the left and right columns.  In the first row, we show the ratio between stellar radius $R_*$ to RL radius, $R_{\rm RL}$, ratio between stellar radius $R_*$ to initial stellar radius $R_{*,i}$, and ratio between RL radius $R_{\rm RL}$ to initial RL radius $R_{\rm RL,i}$.  In the second row, we compare the total mass loss $\Delta M_{\rm tot}$, mass loss due to stellar wind $\Delta M_{\rm wind}$, and mass loss due to RLOF $\Delta M_{\rm RLOF}$ in units of initial stellar mass $M_{*,i}$.  In the third row, we compare the change in stellar spin angular momentum $\Delta J_{\rm spin}$ with the change in orbital angular momentum $\Delta J_{\rm orb}$ in units of $J_*,i = M_{*,i} \sqrt{G M_{*,i} R_{*,i}}$.  { Additionally, we have the solid (dotted) green line as an estimate for the angular momentum of a post-collapse BH mass of $10 M_\odot$ ($5 M_\odot$) and spin of 1.0 (0.5) in units of $J_*,i$}.  In the fourth row, we track the evolution of the helium core mass $M_{\rm HE}$ and carbon-oxygen core mass $M_{\rm CO}$ in units of solar mass $M_\odot$ and central helium-4 mass fraction $M_{^4{\rm He, c}}$ in units of stellar mass $M_*$.  }

    \label{fig:evol21day}
\end{figure*}

\begin{figure*}
    \centering

\includegraphics[width = 0.95\textwidth]{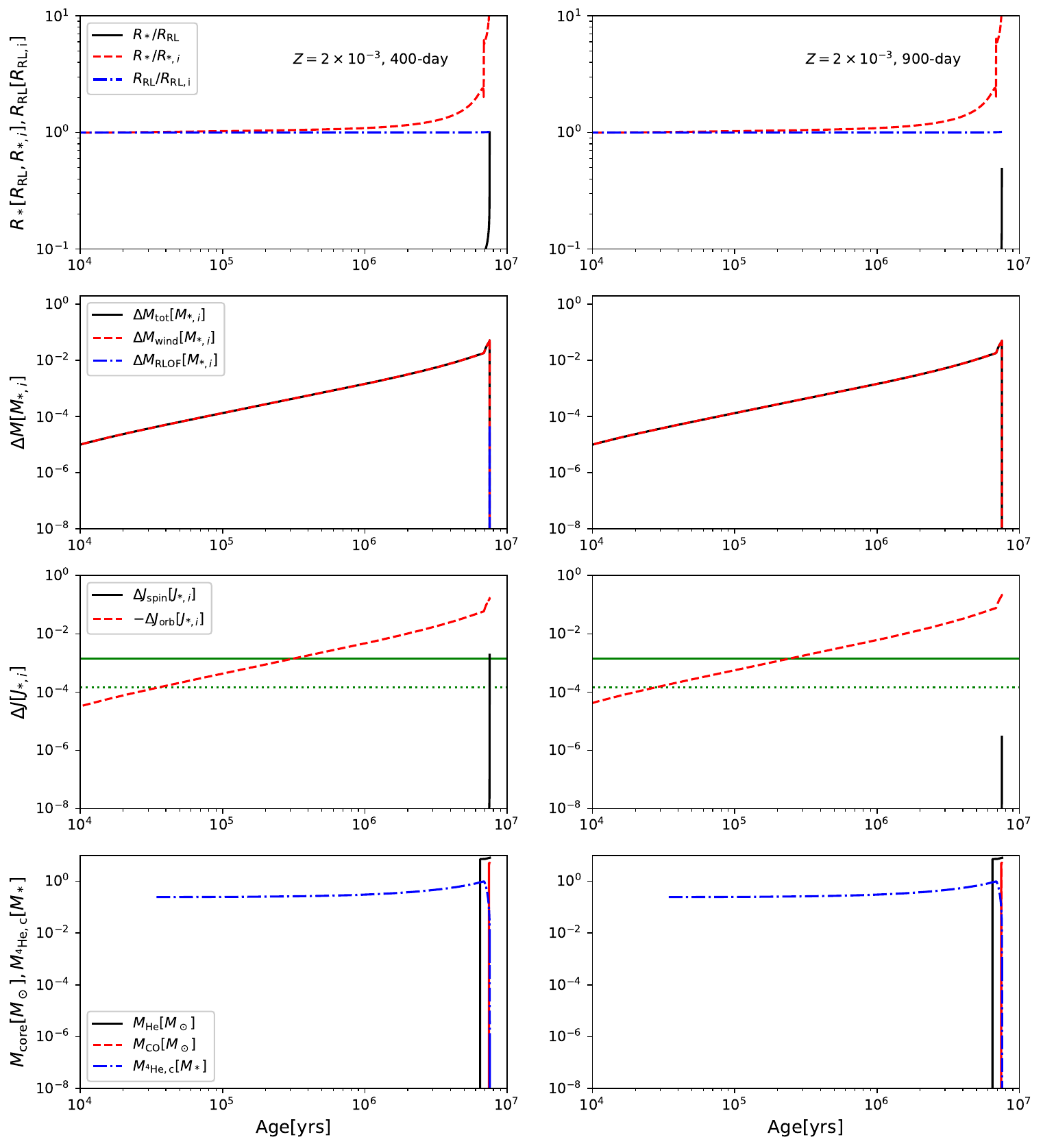}
\caption{{  Focused study of the impact of initial orbital period on mass loss and final stellar spin-up an initially non-rotating MS at stellar termination.  We show the evolution of a MS interacting with a 15$M_\odot$ BH with initial orbital periods at four hundred and nine hundred days in the left and right columns.  Stellar age is given in units of years.
We compare simulation results of a MS of initial mass $M_{*,i}$ = 25$M_\odot$ with metallicity $Z=2 \times 10^{-3}$. }  In the first row, we show the ratio between stellar radius $R_*$ to RL radius, $R_{\rm RL}$, ratio between stellar radius $R_*$ to initial stellar radius $R_{*,i}$, and ratio between RL radius $R_{\rm RL}$ to initial RL radius $R_{\rm RL,i}$.  In the second row, we compare the total mass loss $\Delta M_{\rm tot}$, mass loss due to stellar wind $\Delta M_{\rm wind}$, and mass loss due to RLOF $\Delta M_{\rm RLOF}$ in units of initial stellar mass $M_{*,i}$.  In the third row, we compare the change in stellar spin angular momentum $\Delta J_{\rm spin}$ with the change in orbital angular momentum $\Delta J_{\rm orb}$ in units of $J_*,i = M_{*,i} \sqrt{G M_{*,i} R_{*,i}}$.  { Additionally, we have the solid (dotted) green line as an estimate for the angular momentum of a post-collapse BH mass of $10 M_\odot$ ($5 M_\odot$) and spin of 1.0 (0.5) in units of $J_*,i$}.  In the fourth row, we track the evolution of the helium core mass $M_{\rm HE}$ and carbon-oxygen core mass $M_{\rm CO}$ in units of solar mass $M_\odot$ and central helium-4 mass fraction $M_{^4{\rm He, c}}$ in units of stellar mass $M_*$.  }
    \label{fig:evolZorbit}
\end{figure*}

{ For the binary parameters given in Tab.~\ref{tab:simulation-parameters}, our numerical simulations  over stellar lifetime show a particular region with significant final spin angular momentum of the MS due to the tidal interaction with a BH, consistent with analytic estimates for long GRB binary progenitors \citep{LR22}.}  In this section, we compare our theoretical estimate of the stellar spin-up \(\Delta J_{\rm spin}\) discussed in Sec.~\ref{subsec:theoretical_expectations} 
with \textsc{MESA} simulation results.  In Fig.~\ref{fig:main-results2} and Fig.~\ref{fig:main-results3}, we show the dependence of final stellar spin angular momentum, mass loss, and orbital separation on stellar metallicity and initial orbital period with models given in Tab.~\ref{tab:simulation-parameters} for initially non-rotating MS.  { We present a focused study of initially non-rotating MS in Fig.~\ref{fig:evol21day} and Fig.~\ref{fig:evolZorbit} showing the evolution of stellar radius, RL radius, mass loss due to stellar wind and RLOF, spin and orbital angular momentum, and our termination conditions in tracking the formation of the helium and carbon-oxygen core mass and central helium-4 mass fraction. Additionally, in Fig.~\ref{fig:radprof}, we track the radial evolution in profiles of stellar density and rotation.  In Fig.~\ref{fig:initial_rot1} and Fig.~\ref{fig:initial_rot2}, we provide a focused study on the impact of initial rotation on representative models at short and long initial orbital periods.  For reference, in the right column of Fig.~\ref{fig:main-results2}, we show cyan star markers which correspond to specific simulation parameters shown in Fig.~\ref{fig:evol21day}-\ref{fig:initial_rot2}.  In Fig.~\ref{fig:initial_rot_sample}, we show the final mass loss and spin angular momentum at stellar termination for both initially non-rotating and highly rotating MS.  }
In the following, we describe our results in detail with regards to the identification of the parameter space of GRB progenitor binary systems, mass loss in late stage evolution due to RLOF, angular momentum, termination conditions, metallicity, mass-ratio, rotation, and additional tests and caveats.

%%%%%%%%%%%%%%%%%%%%%%%%%%%%%%%%%%%%%%%%%%%%%%%%%%%%%%%%%%

%%%%%%%%%%%%%%%%%%%%%%%%%%%%%%%%%%%%%%%%%%%%%%%%%%%%%%%%%%

%%%%%%%%%%%%%%%%%%%%%%%%%%%%%%%%%%%%%%%%%%%%

%%%%%%%%%%%%%%%%%%%%%%%%%%%%%%%%%%%%%%%%%%%%%%%%

\subsection{Progenitor binary systems with significant stellar spin-up and negligible mass loss}

{  We identify a region of parameter space where the tidal interaction in the binary leads to significant stellar spin angular momentum such that after the MS collapses, its remnant BH-disk system may retain enough angular momentum to launch a jet via Blandford-Znajek process \citep{BZ77}.  }

The angular momentum of the remnant BH of mass $M$ and spin $a$ may be estimated as $a(GM^2/c)$ from dimensional analysis and we use this as the viable progenitor threshold to compare with our simulations.  { For a GRB to be created from such a progenitor binary system, we estimate the lower limit for the final spin angular momentum of $\gtrsim 10^{50}$ g$\cdot$cm$^{2}$s$^{-1}$ using a BH mass range of $5-10 M_\odot$ and BH spin range of 0.5-1.0.  In the top rows of Fig.~\ref{fig:main-results2} and Fig.~\ref{fig:main-results3}, we show the \textsc{MESA} simulation results of the  final stellar spin-up at stellar termination for initially non-rotating MS and compare with the lower limit range of possible GRB progenitors.  Specifically, the maximum value of the final spin-up for all simulations in the case of the $25 M_\odot$ MS interacting with a $10 M_\odot$ ($15 M_\odot$) is $2.63 \times 10^{52}$ g$\cdot$cm$^{2}$s$^{-1}$ ($2.48 \times 10^{52}$ g$\cdot$cm$^{2}$s$^{-1}$).  The maximum value of the final spin-up for all simulations in the case of the $15 M_\odot$ MS interacting with a $10 M_\odot$ ($15 M_\odot$) is $1.47 \times 10^{51}$ g$\cdot$cm$^{2}$s$^{-1}$ ($3.67 \times 10^{52}$ g$\cdot$cm$^{2}$s$^{-1}$). 
In Fig.~\ref{fig:main-results2} and Fig.~\ref{fig:main-results3}, we identify the region of parameter space of possible GRB progenitors as inside the solid (dotted) contour line that is green for the top and middle panels and white for the bottom panel.  This line approximately represents the value $10^{51} (10^{50}) {\rm g}\cdot{\rm cm}^2 {\rm s}^{-1}$, the estimate for the angular momentum of a post-collapse BH mass of $10 M_\odot$ ($5 M_\odot$) and spin of 1.0 (0.5).  }

%We have the solid (dotted) green line as an estimate for the angular momentum of a post-collapse BH mass of $10 M_\odot$ ($5 M_\odot$) and spin of 1.0 (0.5).  
 
{ Across the parameter space in Fig.~\ref{fig:main-results2} and Fig.~\ref{fig:main-results3}, the simulation results for the final spin-up \(\Delta J_{\rm spin}\) of an initially non-rotating MS are significant across a roughly diagonal curve through parameter space beginning at shorter initial orbital periods and lower metallicities and ending at longer initial orbital periods and higher metallicities.  We find that the lower limit of possible GRB progenitors is exceeded for regions with similar morphology to Fig.~\ref{fig:main-estimate} identified by a ridge of maximum $\Delta J_{\rm spin}$ across all the simulations which traces the limit of RLOF.  }  In comparing the top rows with the middle rows of Fig.~\ref{fig:main-results2} and Fig.~\ref{fig:main-results3}, we find that the boundary of the ridge coincides with the onset of appreciable mass loss from RLOF (and indicated by the estimate in Fig.~\ref{fig:main-estimate}).  { We use for reference, in all panels, a dashed cyan line as an approximate contour that indicates a mass loss of 10\% of the initial mass of the star as the boundary of significant mass loss.  }
Furthermore, in comparing the top rows with the bottom rows, for potential progenitor systems, the binary interactions lead to mostly negligible changes to the orbital separations ($\sim 1\%$) and $\lesssim 10\%$ smaller for limited regions associated with the RLOF boundary.  Otherwise, the final orbital separations are larger by a factor of $\lesssim 10$.  

{ We now inspect specific cases in relation to the diagonal curve through the parameter space that gives rise to a final spin-up above the lower limit of possible GRB progenitors.  Simulation results at the lowest metallicity \(Z\sim 10^{-4}\) give rise to significant final stellar spin-ups at initial orbital periods \(\tau_{\rm orb, i}\approx20\)–\(50\) days associated with negligible mass loss \(\Delta M_{\rm tot}/M_{*,i}\lesssim10^{-2}\) and changes to the orbital separations \(a_{\rm orb}/a_{\rm orb,i}\sim 1\).  At these relatively short initial orbital periods, a higher metallicity \(Z \sim 10^{-3}\) MS will lead to regions approaching RLOF \(R_{*,f}/R_{\rm RL}\sim1\), where the MS will lose more mass, and the final stellar spin-up will fall below the progenitor threshold.  Alternatively, at higher metallicities \(Z \sim 10^{-3}\) and larger initial orbital separations \(\tau_{\rm orb, i}\approx10^2\)–\(5\times 10^2\), the final stellar spin-up is above this threshold and the tidal interactions also give rise to negligible mass loss \(\Delta M_{\rm tot}/M_{*,i}\lesssim10^{-2}\) and changes to the orbital separations \(a_{\rm orb}/a_{\rm orb,i}\sim 1\).   At the highest metallicities considered in this study ($Z \sim 4 \times 10^{-3}-10^{-2}$), the final stellar spin-up is above the threshold for very long initial orbital separations of $\tau_{\rm orb,i} \approx {2-4}\times 10^3$ days whose tidal interactions lead to negligible mass loss \(\Delta M_{\rm tot}/M_{*,i}\lesssim10^{-2}\) and changes to the orbital separations \(a_{\rm orb}/a_{\rm orb,i}\sim 1\).  Otherwise, the final stellar spin up is below the threshold and is associated with significant mass loss \(\Delta M_{\rm tot}/M_{*,i}\gtrsim10^{-1}\) and moderate changes to the orbital separations \(a_{\rm orb}/a_{\rm orb,i}\lesssim 10\).   }

These trends align with the work by \citet{Det08}, where it is shown that, while the spin-up process itself may work, significant mass loss due to winds and RLOF negate the spin-up effect through the corresponding loss in stellar angular momentum.  Additionally, our results are comparable to the orbital separation to tidal radius and mass ratio regime described in the parameter study simulating the interaction between stellar BH and main-sequence stars in \citet{Kremer2022}.   We note that inhomogeneities arise in the final stellar spin-up parameter space map of Fig.~\ref{fig:main-results2} and Fig.~\ref{fig:main-results3} and this is likely due to the choice in the resolution in initial orbital period and metallicity for these simulations.  Our goal in running the sample of \textsc{MESA} simulations in Tab.~\ref{tab:simulation-parameters} is to survey the landscape which give rise to GRB progenitors.  We expect follow-up investigations at higher resolution will provide higher accuracy in the mapping of the parameter space consistent within the morphology.

{ Finally, we note a small region of parameter space for the initially non-rotating simulations at high metallicity ($\sim 10^{-2}$) and short initial orbital periods (top left of contours in Fig.~\ref{fig:main-results2} and Fig.~\ref{fig:main-results3}) showing a notable increase in spin angular momentum when significant mass transfer occurs.  We interpret these results as reliable given our resolution studies discussed in Appendix Sec.~\ref{sec:convergence} at greater mass loss with much shorter initial orbital periods at the wide range of metallicities.  As we will discuss further in the results below, for this region of parameter space, the tidal process removes considerable mass from the MS, leaving a rotating core with significant spin for a limited range of short initial orbital period.
}

\subsection{Roche-lobe overflow in late stage stellar evolution}
\label{subsec:Rochelobe}
  For all simulations in Tab.~\ref{tab:simulation-parameters}, the radius of the MS increases significantly towards the late-stages of stellar evolution in both single star (as discussed in Appendix Sec.~\ref{sec:appendix_singlestar}) and binary models.  While the RL radius remains relatively constant, there is an increase in the ratio between stellar to RL radius, leading to significant mass loss.  { The focused studies of the impact of stellar metallicity in Fig.~\ref{fig:evol21day} and initial orbital period in Fig.~\ref{fig:evolZorbit} on mass loss show that the primary drivers of mass loss are due to stellar winds and RLOF in binary simulations of the interaction between a 25$M_\odot$ MS and a 15$M_\odot$ BH.}  In the first row of both figures, we compare the ratio of stellar to RL radius $R_*/R_{\rm RL}$ (solid line), stellar to initial stellar radius $R_*/R_{*,i}$ (dashed line), and Roche flow overflow to initial Roche flow overflow radius $R_{\rm RL}/R_{\rm RL, i}$ (dashed dotted line) over stellar lifetime.  
{ We find that the time of increase in stellar radius close to termination is coincident with a sharp increase in total mass loss $\Delta M_{\rm tot}$, given in the second row.  Additionally, in these studies, we compare the evolution of total mass loss $\Delta M_{\rm tot}$, mass loss due to stellar wind $\Delta M_{\rm wind}$, and mass loss due to RLOF $\Delta M_{\rm RLOF}$ in units of initial stellar mass $M_{*,i}$.  The results given in Fig.~\ref{fig:evol21day} and Fig.~\ref{fig:evolZorbit} are representative of all of the simulations listed in Tab.~\ref{tab:simulation-parameters} where mass loss is driven by stellar winds prior to the sharp increase in stellar radius (at less than 
$\sim 7$ Myr).  Mass loss is potentially driven by RLOF after the rapid radial increase and whether or not it is the driving mechanism depends on stellar metallicity and initial orbital period.  In Fig.~\ref{fig:evol21day} ($\tau_{\rm orb, i} = 21$ days, $Z=10^{-4}, 10^{-2}$), we show the impact of low (left column) and high (right column) stellar metallicity in comparing binary interactions at the same initial twenty-one day orbital period.  }In comparing both columns of the second row, we see that at higher metallicity ($Z=10^{-2}$) RLOF is the main driver of mass loss whereas for lower metallicity ($Z=10^{-4}$) its contribution is less than the mass loss due to stellar wind.  { In Fig.~\ref{fig:evolZorbit} ($\tau_{\rm orb, i} = 400, 900$ days, $Z=2\times 10^{-3}$), we show the impact of initial orbital period at four hundred days (left column) and nine hundred days (right column) in comparing binary interactions with MS at the same stellar metallicity of $Z=2\times 10^{-3}$.}  In the top row, at the end of stellar evolution, the ratio of stellar radius to RL radius is unity for the shorter initial orbital period and less than unity for the longer initial orbital period.  In the second panel, we find that in both cases, the dominant mechanism for mass loss is due to stellar winds where the mass loss due to RLOF is larger in the left column than the right column.  For both Fig.~\ref{fig:evol21day} and Fig.~\ref{fig:evolZorbit}, we note dips in the late-stage evolution of the expansion in stellar radius for simulations with significant mass loss and suspect this may be related to the numerical challenges in modeling the stellar envelope in a 1D stellar evolution code discussed in \citet{Paxton2013}.

%%%%%%%%%%%%%%%%%%%%%
\begin{figure*}
   \centering
\includegraphics[width = 0.95\textwidth]{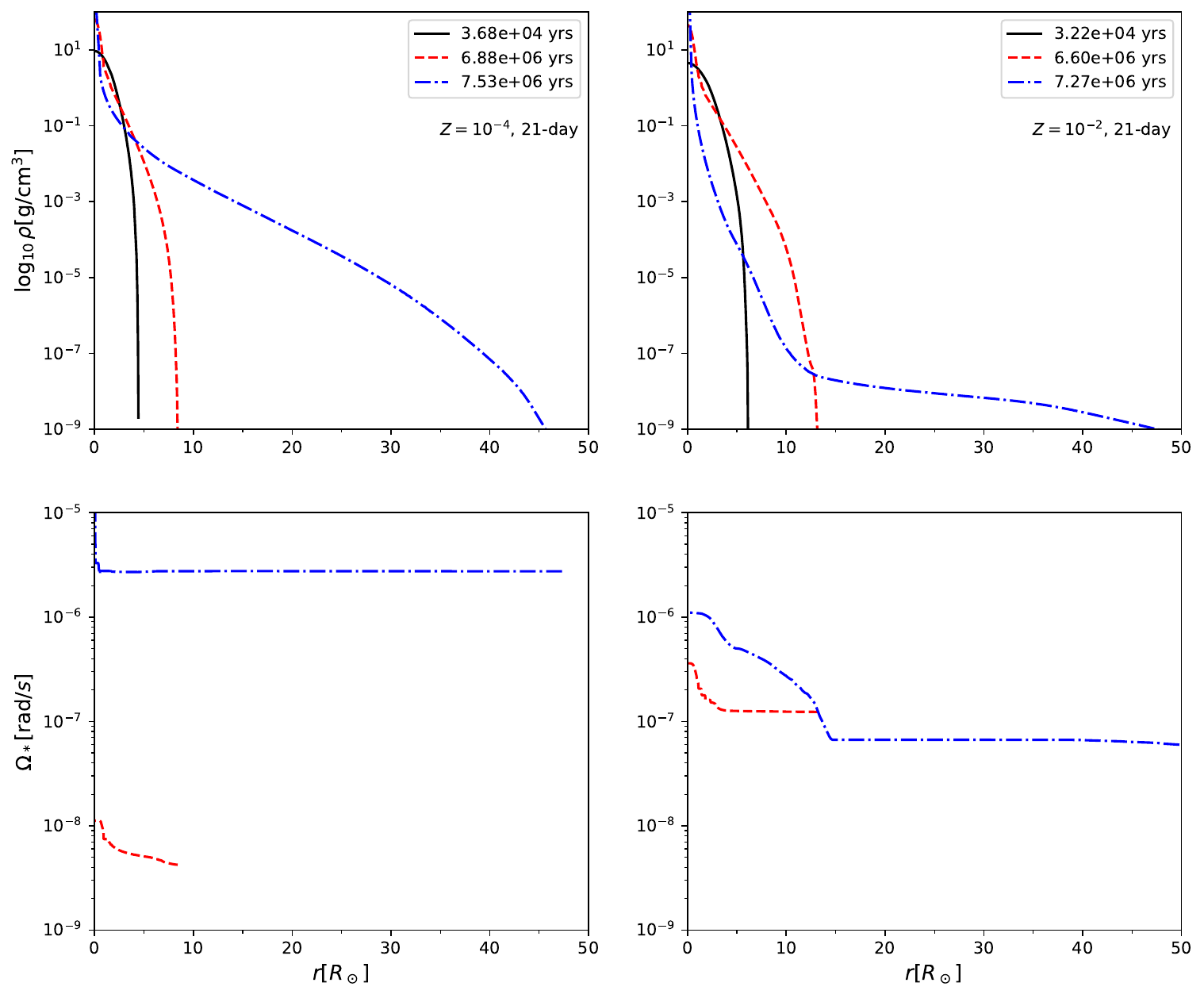}
\caption{  { Impact of stellar metallicity on the radial evolution of an initially non-rotating MS in stellar density and rotation.  } We compare profiles for a 25 $M_{\odot}$ star at low metallicity ($Z=10^{-4}$) in the left panel and high metallicity ($Z=10^{-2}$) in the right panel at an initial orbital period of twenty-one days with a $15 M_\odot$ BH at three stellar ages as indicated by the legend.  Radius is given in units of solar radius $R_\odot$.  The top row shows the evolution of stellar density $\rho$ in CGS units.  The bottom row shows the corresponding profile of stellar rotation $\Omega_*$ in radians per second.  Note that the stellar rotation profiles at early stages of the evolution is close to numerical round-off error $\lesssim 10^{-16}$ because the star is initially non-rotating.
}
    \label{fig:radprof}
\end{figure*}
\subsection{Angular momentum}

{ We now discuss the limit in the parameter space which gives rise to either a significant increase or increase then diminishment in the final spin angular momentum of the MS.   The focused studies for initially non-rotating MS on the impact of stellar metallicity in Fig.~\ref{fig:evol21day} and initial orbital period in Fig.~\ref{fig:evolZorbit} on final stellar spin-up show, in the third row,
 the change in spin angular momentum $\Delta J_{\rm spin}$ (solid black line) in units of stellar angular momentum at break-up velocity $J_{*,i}$, where $J_{*,i} = M_{*,i} \sqrt{G M_{*,i} R_{*,i}}$.  Additionally, we show a solid (dotted) green line as an estimate for the angular momentum of a post-collapse BH mass of $10 M_\odot$ ($5 M_\odot$) and spin of 1.0 (0.5) in units of $J_*,i$.  For all of the simulations, the spin-up is characterized by an increase coincident with the increase in stellar radius.  The behavior of simulations in Tab.~\ref{tab:simulation-parameters} that are not subjected to significant mass loss are well represented by the left column of Fig.~\ref{fig:evol21day} and both columns of Fig.~\ref{fig:evolZorbit} where total mass loss $\Delta M_{\rm tot}$ is less than 10\% of the initial stellar mass and the final spin angular momentum is greater than zero.  In comparing the columns of Fig.~\ref{fig:evolZorbit} at fixed stellar metallicity ($Z=2 \times 10^{-3}$), we find that significant stellar spin-up above the progenitor threshold of the green solid (dotted) line depends on the strength of the tidal interaction determined by the initial orbital period.  For the relatively low, four hundred day initial orbital period, we find that the final stellar spin up is above this threshold.  For the relatively high, nine hundred day initial orbital period, we find that the final stellar spin up is well below this threshold.  The behavior of simulations in Tab.~\ref{tab:simulation-parameters} that are subjected to significant mass loss are well represented by the right column of Fig.~\ref{fig:evol21day} where total mass loss $\Delta M_{\rm tot}$ is greater than 10\% of the initial stellar mass and with significantly low final spin angular momentum.  For these cases, we find that while there is an increase in stellar spin-up that corresponds with the increase in stellar radius, there is also a sharp diminishment after the maximum peak coincident with mass loss due to significant RLOF.  }

{ Notably, we find a region in parameter space for initially non-rotating MS which gives rise to a significant spin-up by $\sim 10^{-2} J_{*,i}$, above the limit of possible GRB progenitors, while the tidal torque and mass loss lead to slight decreases in the orbital angular momentum.  This finding is consistent with the small changes ($\lesssim 10\%$) in orbital separation shown in Fig.~\ref{fig:main-results2} and Fig.~\ref{fig:main-results3}.  In the third row of the focused studies in Fig.~\ref{fig:evol21day} and Fig.~\ref{fig:evolZorbit}, we show the change in spin angular momentum $\Delta J_{\rm spin}$ and the corresponding change in orbital angular momentum $\Delta J_{\rm orb}$ of the MS in units of $J_{*,i}$.  In particular, at low metallicity $Z=10^{-4}$ given in the left column of Fig.~\ref{fig:evol21day}, the first, second, and third rows show that the star only fills its Roche lobe at the very end of the evolution, so the brief RLOF event removes much less mass than the stellar wind, yet the stellar spin still increases significantly.  The orbital angular momentum decreases slightly, matching the change in spin angular momentum increase.  For higher metallicity ($Z=2 \times 10^{-3}$) and greater wind-driven mass loss, given in the left column of Fig.~\ref{fig:evolZorbit}, we find a lower magnitude in the peak value of spin angular momentum and a larger decrease in orbital angular momentum that does not match the increase in spin angular momentum.  Overall, we find that for MS -- BH interactions with negligible mass loss, the decrease of $\Delta J_{\rm orb}$ matches the increase in $\Delta J_{\rm spin}$. For binaries with substantial mass loss, especially when the MS undergoes RLOF, the sharp drops in $\Delta J_{\rm orb}$ coincide with the mass loss events but are much larger than the simultaneous gains in $\Delta J_{\rm spin}$. This shows that a significant fraction of the orbital angular momentum is carried away by the escaping mass (either due to RLOF or stellar winds especially at higher metallicity), rather than being transferred into the spin of the MS.
}

The values of the final spin angular momentum of the MS for simulations in Fig.~\ref{fig:main-results2} and Fig.~\ref{fig:main-results3} are given in CGS units.  The corresponding values in final specific angular momentum range from $\sim 3 \times 10^{15}\  {\rm cm^{2} s^{-1}}$ to $5 \times 10^{17}\ {\rm cm^{2} s^{-1}}$.  This has important implications for the spin of the BH remnant when this MS collapses, as well as the angular momentum in the disk that forms around the BH during/just after collapse.  We consider this in our Discussion section below.

\subsection{Dependence on Metallicity}
\label{subsec:metallicity}
We now discuss our results in terms of  stellar metallicity within the low ($Z=10^{-4}$) and high ($Z=10^{-2}$) range of parameter space.   { In Fig.~\ref{fig:main-results2} and Fig.~\ref{fig:main-results3}, we find contributions to the parameter space of progenitors corresponding to initial orbital periods of $\lesssim 500$ days at low metallicity with the smallest at $\sim 20$ days at $Z\sim 10^{-4}$ and $\lesssim 500$ days at $Z\sim 10^{-3}$.  Additionally, at high metallicity we find contributions at larger initial orbital periods of $2-4 \times 10^{3}$ days at $Z \sim 4 \times 10^{-3} - 10^{-2}$.}  The shape of the region follows the boundary of significant or negligible mass loss where the stellar radius to RL radius ratio is close to unity, $R_{*,f}/R_{\rm RL} \sim 1$.  This is primarily the case because of two reasons related to the discussion in Appendix Sec.~\ref{sec:appendix_singlestar} and Sec.~\ref{subsec:Rochelobe}.  First, within the lifetime of a MS, the stellar radius increases with increasing metallicity and therefore MS with larger metallicities are subject to stronger tidal interactions for a given orbital period than MS with smaller metallicities.  Furthermore, when the stellar radius increases at the late stages of evolution, higher metallicity MS will be subject to greater RLOF.  Second, the mass loss associated with stellar winds increases with stellar metallicity.  This combined effect leads to significant mass loss for high metallicity stars and diminished spin angular momentum below the lower limit for GRB progenitors.  { In Fig.~\ref{fig:evol21day}, we compare results between low and high metallicity stars for twenty-one day orbital periods.  } In the first row, we find the stellar to RL radius, $R_*/R_{\rm RL}$, is closer to unity for high metallicity MS throughout its lifetime.  In the second row, we see a corresponding greater mass loss due to both stellar wind and RLOF.  In the third row, we find that this leads to greater diminishment in spin angular momentum.

{ We consider the impact of stellar metallicity in the evolution of an initially non-rotating MS in radial profiles of stellar density and rotation.  In Fig.~\ref{fig:radprof}, we compare profiles at early, middle, and late-stages of stellar evolution for a 25 $M_{\odot}$ star at low metallicity ($Z=10^{-4}$)  and high metallicity ($Z=10^{-2}$) at an initial orbital period of twenty-one days with a $15 M_\odot$ BH.  In the top row, stellar density $\rho$ is given in CGS units versus stellar radius, given in solar radius $R_\odot$.  The bottom row shows the corresponding profile of stellar rotation $\Omega_*$ in radians per second.  At early stages, given by the solid line at stellar ages $3.2-3.7 \times 10^4$ yrs, the stellar rotation profiles are close to numerical round-off error $\lesssim 10^{-16}$ because the star is initially non-rotating.  At middle stages, given by the dashed line at at $6.6-6.9 \times 10^{6}$ yrs, we find that the MS has expanded in radius close to a factor of two in comparison to early stages.  At late stages, given by the dashed dotted line at at $7.3-7.5 \times 10^{6}$ yrs, we find that the MS has significantly expanded in radius close to a factor of ten.  This is consistent with the evolution of stellar radius in Fig.~\ref{fig:evol21day}, where significant mass loss through RLOF does not occur for the low metallicity case, but RLOF does occur at late stages for the high metallicity case.  Nonetheless, in both cases, we identify a late time rise in stellar rotation (blue dot dashed curve in the bottom panels) that coincides with the increase in stellar radius and resulting from the appreciable tidal torque.  Furthermore, we note that stellar rotation for the low metallicity case is uniform across the radial profile of the star.  For the high metallicity case, where significant mass loss has occurred at late stages, we find a significant drop in density especially in the outer layers of the star suggesting that the tidal process removes mass from the envelope.  As a result, the radial profile of stellar rotation shows higher rotation at the remnant core than the outer layers.
}
\subsection{Dependence on Stellar and BH Masses}

{  We show the \textsc{MESA} simulation results of binary interactions between MS of initial masses $M_{*,i} [M_\odot] = \{15, 25\}$ and BH of initial masses $M_{\rm BH} [M_\odot] = \{10, 15\}$ which correspond to comparable mass ratios, $M_*/M_{\rm BH} = \{1.0, 1.5, 1.667, 2.5\}$.}  We find consistency with the theoretical estimate discussed in Sec.~\ref{subsec:theoretical_expectations}.  { In Fig.~\ref{fig:main-results2} and Fig.~\ref{fig:main-results3}, we show that the range in parameter space of progenitors is similar in morphology across all binary pairs of MS and BH masses.  As expected, the location of the RLOF limit has the most significant dependence on the chosen MS initial mass, rather than BH initial mass.  We find, consistent with the differences between the left and right columns of the theoretical expectations of Fig.~\ref{fig:main-estimate}, that the boundary of significant mass loss is primarily shifted from the $25 M_\odot$ MS case of Fig.~\ref{fig:main-results2} to longer initial orbits and lower metallicities to the $15 M_\odot$ MS case of Fig.~\ref{fig:main-results3}.  Additionally, in the former case, the boundary gradually reaches the high metallicity limit at $Z=10^{-2}$ at an initial orbital period of $\sim 3\times 10^3$ days, but in the latter case, the boundary sharply cuts off between metallicity ranges $2\times 10^{-3} - 10^{-2}$ at an initial orbital period of $\sim 10^3$ days.  

The competing effects of significant tidal torque and mass loss lead to two distinct branches, similar in morphology across all binary pairs of MS and BH masses, at relatively shorter initial orbital periods $\lesssim 10^3$ days and longer initial orbital periods $>10^3$ days.  For the first branch, the magnitude of final stellar spin-up $\Delta J_{\rm spin}$ and range in parameter space is greater and more expansive, respectively, corresponding to a MS initial mass of $25M_\odot$ (Fig.~\ref{fig:main-results2}) than a MS initial mass of $15M_\odot$ (Fig.~\ref{fig:main-results3}).  Additionally find this branch is shifted towards lower metallicity $\lesssim 10^{-3}$ for the MS initial mass of $15M_\odot$.  For a given MS initial mass, the region of significant stellar spin up is comparable in magnitude of $\Delta J_{\rm spin}$ and more expansive for the larger BH mass.  Inspecting the parameter space region that gives rise to spin-ups above the progenitor thresholds, we find that for initial orbital periods less than $10^3$ days, binary systems with a $15M_\odot$ MS are less likely $25M_\odot$ MS than to form 10$M_\odot$ post-collapse BH (green solid line).  Additionally, this region of potential progenitors is shifted towards the parameter space of lower metallicity $\lesssim 10^{-3}$ for the $15M_\odot$ MS case.  For the second branch, at longer initial orbital periods $>10^3$ days, we find another region of negligible mass loss and significant final spin up at high metallicity $\gtrsim 4\times 10^{-3}$.  For all binary pairs of MS and BH masses, we have a small region of spin-ups above the lower estimate for a post-collapse BH progenitor at 5$M_\odot$ at high metallicity $\lesssim 10^{-2}$ (green dotted line).  Notably, for the binary system of a $15M_\odot$ MS and $15M_\odot$ BH, we have a very high spin-up above the higher estimate (green solid line) that results from the "sweet spot" of maximum tidal torque and minimum mass loss.
}

%%%%%%%%%%%%%%%%%%%%%
\begin{figure*}
   \centering
\includegraphics[width = 0.95\textwidth]{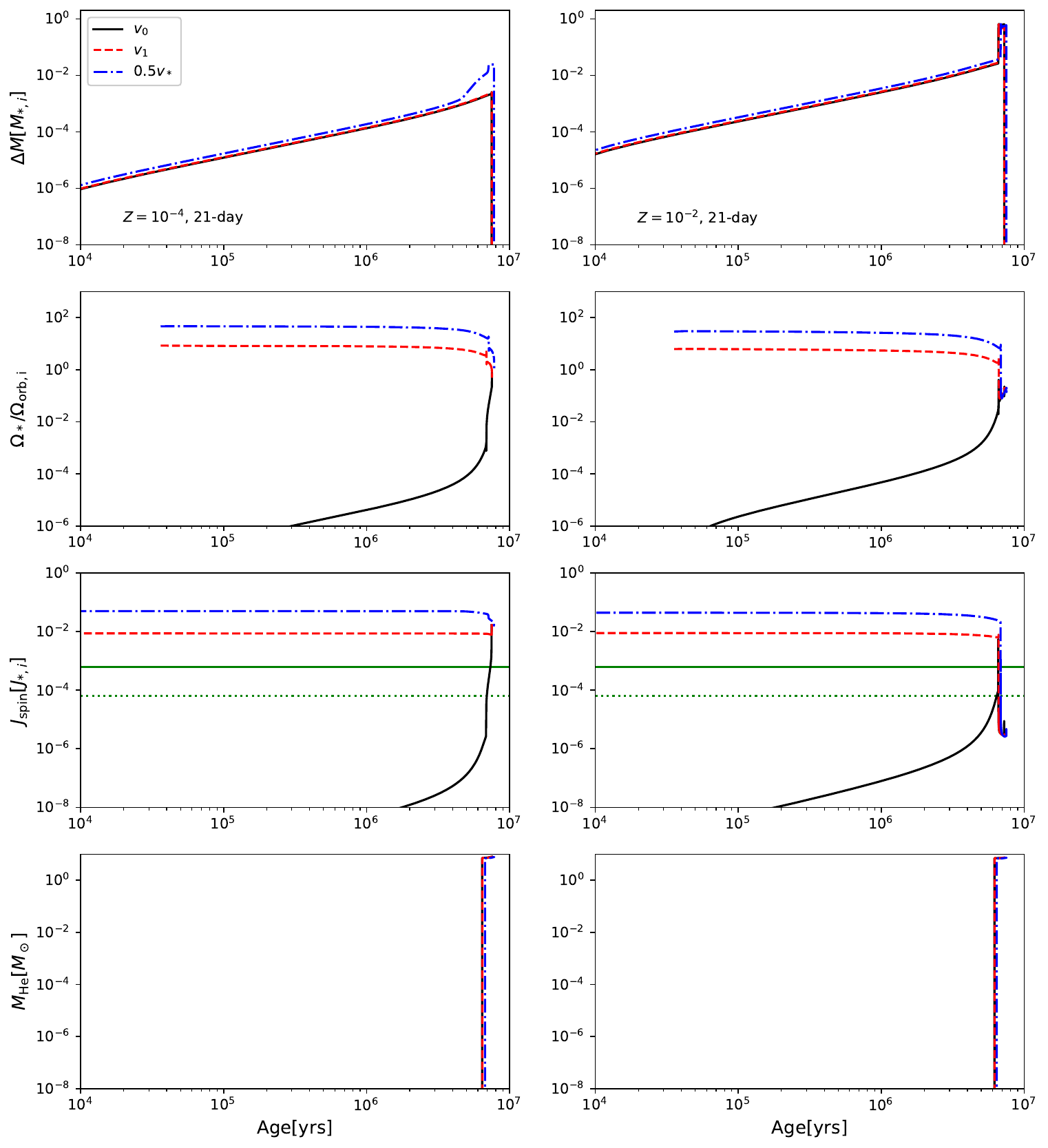}
\caption{{ Focused study of the impact of initial stellar rotation for binaries that are tidally synchronized.}
We compare simulation results initialized at three stellar velocities ($v_0$, $v_1$, $0.5v_*$) under an estimate of the critical limit of stellar break-up of a MS of initial mass $M_{*,i}$ = 25$M_\odot$ with metallicity $Z=10^{-4}$ and $Z=10^{-2}$ in the left and right columns.  The mass of the BH is 15$M_\odot$ and stellar age is given in units of years.  { The initial orbital period is twenty-one days.}  In the first row, we compare the total mass loss $\Delta M$.  In the second row, we compare the ratio of stellar rotational velocity $\Omega_*$ to initial orbital velocity $\Omega_{\rm orb, i}$.  In the third row, we compare the spin angular momentum $J_{\rm spin}$, in units of $J_*,i = M_{*,i} \sqrt{G M_{*,i} R_x`{*,i}}$.  
{ Additionally, we have the solid (dotted) green line as an estimate for the angular momentum of a post-collapse BH mass of $10 M_\odot$ ($5 M_\odot$) and spin of 1.0 (0.5) in units of $J_*,i$}.
In the fourth row, we track the evolution of the helium core mass $M_{\rm HE}$ in units of solar mass $M_\odot$.  
}
    \label{fig:initial_rot1}
\end{figure*}

%%%%%%%%%%%%%%%%%%%%%
\begin{figure*}
   \centering
\includegraphics[width = 0.95\textwidth]{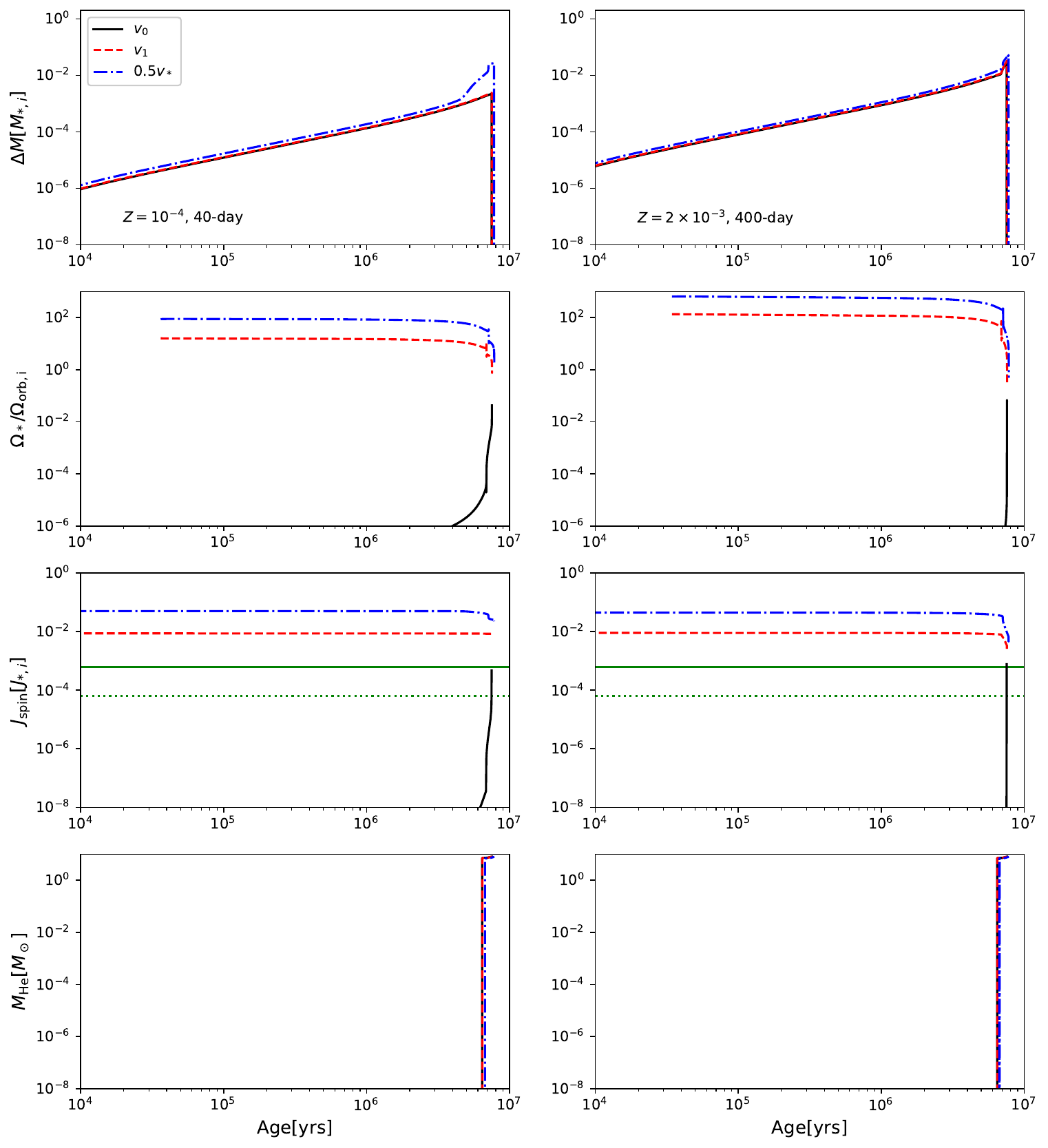}
\caption{  { Focused study of the impact of initial stellar rotation for binaries that are not tidally synchronized.}
We compare simulation results initialized at three stellar velocities ($v_0$, $v_1$, $0.5v_*$) under an estimate of the critical limit of stellar break-up of a MS of initial mass $M_{*,i}$ = 25$M_\odot$.  In the left column, we show results at metallicity $Z=10^{-4}$ at a forty day initial orbital period.  In the right column, we show results at metallicity $Z=2 \times 10^{-3}$ at a four hundred day initial orbital period.  The mass of the BH is 15$M_\odot$ and stellar age is given in units of years.  In the first row, we compare the total mass loss $\Delta M$.  In the second row, we compare the ratio of stellar rotational velocity $\Omega_*$ to initial orbital velocity $\Omega_{\rm orb, i}$.  In the third row, we compare the spin angular momentum $J_{\rm spin}$, in units of $J_*,i = M_{*,i} \sqrt{G M_{*,i} R_{*,i}}$.  { Additionally, we have the solid (dotted) green line as an estimate for the angular momentum of a post-collapse BH mass of $10 M_\odot$ ($5 M_\odot$) and spin of 1.0 (0.5) in units of $J_*,i$}.In the fourth row, we track the evolution of the helium core mass $M_{\rm HE}$ in units of solar mass $M_\odot$.}
    \label{fig:initial_rot2}
\end{figure*}

 %%%%%%%%%%%%%%%%%%%%%
\begin{figure*}
   \centering
\includegraphics[width = 0.95\textwidth]{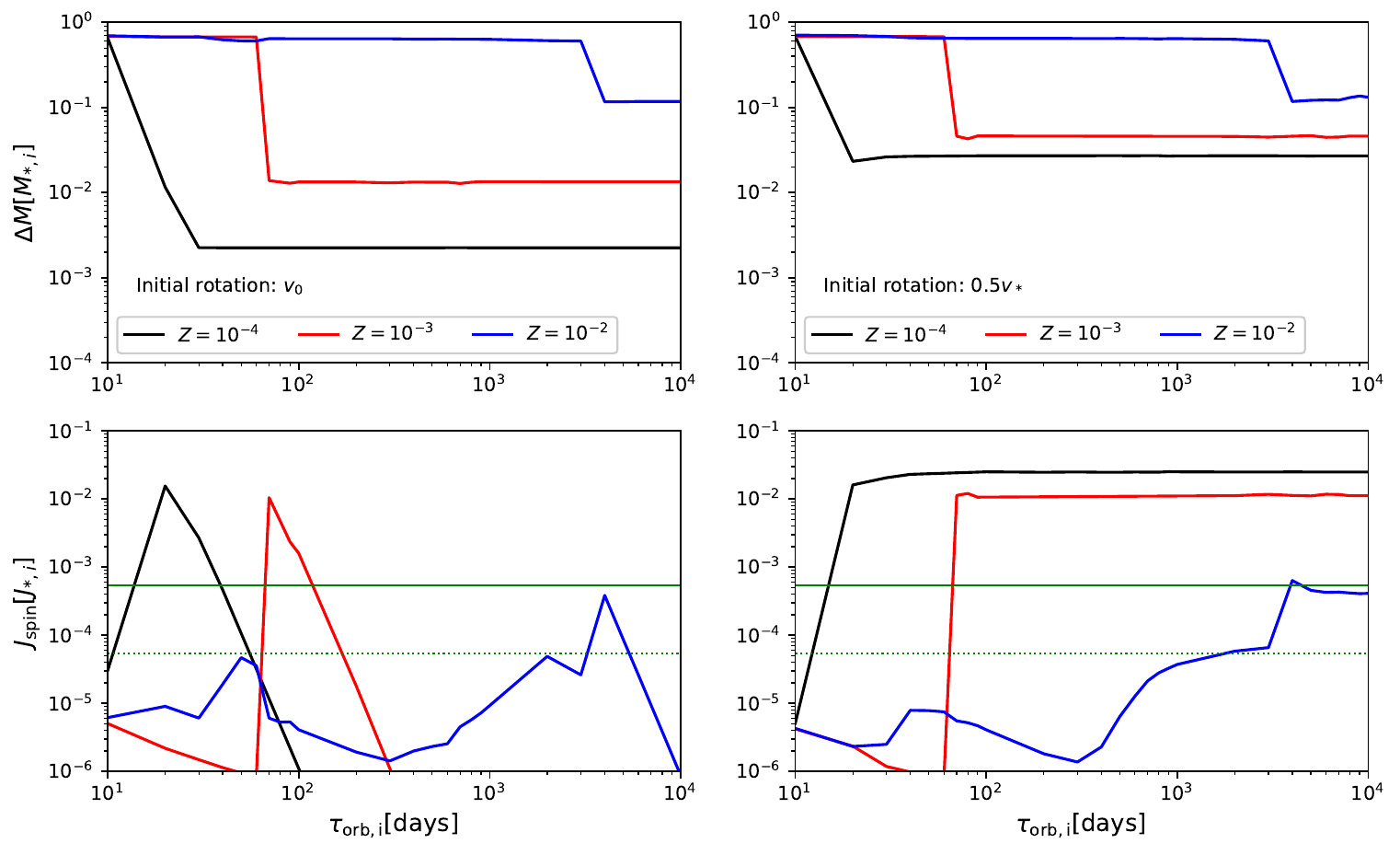}
\caption{  { Binary interactions with negligible mass loss sustain highly spinning MS above the GRB progenitor threshold limit.  We show the final mass loss and spin-up for an initially non-rotating and highly rotating 25 $M_\odot$ MS at a range of stellar metallicities and initial orbital periods with a 15 $M_\odot$ BH.  
 The final total mass loss $\Delta M$ is given in units of $M_{*,i}$ and total spin angular momentum $J_{\rm spin}$ is given in units of $J_{*,i}$.  Each simulation is initialized at orbital period $\tau_{\rm orb, i}$ in days.  Simulations with initial rotations at zero ($v_0$) are given in the left panel and approximately half the break-up velocity ($0.5 v_*$) are given in the right panel for metallicities $Z=10^{-4}$ (black), $Z=10^{-3}$ (red), and $Z=10^{-2}$ (blue).  Additionally, we have the solid (dotted) green line as an estimate for the angular momentum of a post-collapse BH mass of $10 M_\odot$ ($5 M_\odot$) and spin of 1.0 (0.5) in units of $J_*,i$}.}
    \label{fig:initial_rot_sample}
\end{figure*}

%%%%%%%%%%%%%%%%%%%%%%%%%%%%%%%%%%%%%%%%%%%%

\subsection{Dependence on Initial Stellar Rotation}
     
 { We consider the effects of initial stellar rotation on the evolution of a MS in MS-BH binaries that are and are not tidally synchronized.  We compare simulation results of the interaction between a 15 $M_\odot$ BH and a 25 $M_\odot$ MS initialized at three stellar velocities ($v_0$, $v_1$, $0.5v_*$) under an estimate of the critical limit of stellar break-up of a MS.  We note that the initial spin angular momentum values associated with initial stellar velocities $v_1$ and $0.5v_*$ are significantly higher than the highest final spin angular momentum (spin-up) values associated with initially non-rotating MS.  In the left ($\tau_{\rm orb, i} = 21$ days, $Z=10^{-4}$) and right ($\tau_{\rm orb, i} = 21$ days, $Z=10^{-2}$) column of Fig.~\ref{fig:initial_rot1} and the left ($\tau_{\rm orb, i} = 40$ days, $Z=10^{-4}$) and right ($\tau_{\rm orb, i} = 400$ days, $Z=2\times 10^{-3}$) column of Fig.~\ref{fig:initial_rot2}, we show the MS evolution in mass loss, stellar rotation, spin angular momentum, and helium core mass to track stellar termination.  For reference, in the right column of Fig.~\ref{fig:main-results2}, we show cyan star markers which correspond to these specific simulation parameters.  }  Stellar age is given in units of years.  In the first row, we compare the total mass loss $\Delta M$.  In the second row, we compare the ratio of stellar rotational velocity $\Omega_*$ to initial orbital velocity $\Omega_{\rm orb, i}$.  In the third row, we compare the change in stellar spin angular momentum $\Delta J_{\rm spin}$ with the change in orbital angular momentum $\Delta J_{\rm orb}$ in units of $J_*,i = M_{*,i} \sqrt{G M_{*,i} R_{*,i}}$.  In the fourth row, we track the evolution of the helium core mass $M_{\rm HE}$ in units of solar mass $M_\odot$.  { 
 For simulations with non-zero initial rotation, we expect constant stellar rotation and spin angular momentum at the initial values throughout the evolution until significant tidal interactions occur towards the end of stellar evolution when the stellar radius rapidly increases significantly.  In all simulations in this focused study with the chosen high initial rotation, we find that as the stellar radius rapidly increases, the initially high spin angular momentum decreases.  This decrease depends on both mass loss and the strength of the tidal interaction.
  In Fig.~\ref{fig:initial_rot1}, for the initial orbital period of twenty-one days, we compare three initial rotations at metallicities $Z=10^{-4}$ (left column) and $Z=10^{-2}$ (right column).  Previously, we identified the parameters that correspond with the left column as those that lead to negligible mass loss and significant spin-up, from initially zero, above the GRB progenitor threshold (green [dotted] line in third row).  In comparing different initial rotations for the lower limit of stellar metallicity, we find similar evolution of stellar mass loss for a MS initialized at zero ($v_0$) and $\sim$100 km/s ($v_1$), but closer to $\sim$1\% at higher initial rotations of $\sim$520 km/s ($0.5v_*$) at approximately half the stellar break-up velocity for a low metallicity MS.  Both the ratio of stellar to initial orbital rotation and spin angular momentum decrease for the initially spinning MS during the stage of rapid increase in stellar radius.  Notably, this decrease for the highest initial spin ($0.5v_*$) is significantly larger and associated with relatively larger mass loss.  For the initially non-rotating case, the tidal interaction is significant enough, with numerous orbits across the final end stage of stellar lifetime, at an orbital period of twenty-one days to lead to synchronization of the stellar rotation to the orbit, where $\Omega_*/\Omega_{\rm orb, i} \sim 1$.  This is also the case for the initially rotating MS simulations.  As a result, the values of the final spin-up closely convergence to the same value for all three cases.  In comparing different initial rotations for the upper limit of stellar metallicity (right column), we find similar evolution of stellar mass loss for a MS including at higher initial rotations of $\sim$445 km/s ($0.5v_*$).  Both the ratio of stellar to initial orbital rotation and spin angular momentum decrease for the initially spinning MS during the stage of rapid increase in stellar radius and associated with significant mass loss.  Similar to the low metallicity study (left column), for all initial rotation cases at this orbital period, the tidal interaction is significant enough, with numerous orbits across the final end stage of stellar lifetime, to lead to synchronization of the stellar rotation to the orbit.
  As a result, the values of the final spin-up closely convergence to the same value for all three cases.

  In Fig.~\ref{fig:initial_rot2}, we compare results of the three initial rotations $(v_0, v_1, 0.5v_*)$ at two sets of previously identified binary parameters that lead to negligible mass loss and significant spin-up, from initially zero, within the GRB progenitor threshold (green [dotted] line in third row).  At initial orbital periods of forty (left column) and four hundred (right column) days, the tidal interaction is significantly lower than the twenty-one day case of Fig.~\ref{fig:initial_rot1}. 
  In comparing different initial rotations, we find similar results in the evolution of stellar mass loss, with greater mass loss for the higher initial rotation ($0.5v_*$) at a forty day initial orbital period and stellar metallicity $Z=10^{4}$ (left column) than at four hundred days at $Z=2 \times 10^{-3}$ (right column).
Also similar to Fig.~\ref{fig:initial_rot1}, both the ratio of stellar to initial orbital rotation and spin angular momentum decrease for the initially spinning MS during the stage of rapid increase in stellar radius.  However, at the initial orbital periods given in Fig.~\ref{fig:initial_rot2}, the tidal interaction is not significant enough, with multiple orbits across the final end stage of stellar lifetime, to lead to synchronization of the stellar rotation to the orbit, where $\Omega_*/\Omega_{\rm orb, i} < 1$ for the initially non-rotating case.   The values of the final spin angular momentum values do not closely convergence and notably the initially rotating cases remain significantly higher above the progenitor limit by 1-2 orders of magnitude.  

We now consider the dependence of final stellar spin on initial stellar rotation in comparing the final mass loss and spin angular momentum at stellar termination for a range of stellar metallicities and initial orbital periods.  In Fig.~\ref{fig:initial_rot_sample}, we show results for an initially non-rotating (left panel) and highly rotating (right panel) 25 $M_\odot$ MS interacting with a 15 $M_\odot$ BH.  We have the solid (dotted) green line as an estimate for the angular momentum of a post-collapse BH mass of $10 M_\odot$ ($5 M_\odot$) and spin of 1.0 (0.5) in units of $J_*,i$.  The colors black, red, and blue indicate simulations with stellar metallicity $10^{-4}$, $10^{-3}$, and $10^{-2}$, respectively.  In the left panel gives the results for the initially non-rotating case ($v_0$).  The right panel gives the
results for the initially highly spinning MS ($0.5 v_*$), which corresponds to an initial spin angular momentum value well above the threshold estimate of a GRB progenitor system.  In the top and bottom row, we give total mass loss $\Delta M$ in units of $M_{*,i}$ and spin angular momentum $J_{\rm spin}$ in units of $J_{*,i}$ as a function of initial orbital period $\tau_{\rm orb, i}$ in days.  For the initially non-rotating case (left panel), we find the previously discussed final spin angular momentum results of diminishment for binary interactions with significant mass loss and low values where the tidal interaction is weak (long orbital periods).  We compare initial stellar rotation results at metallicity $Z=10^{-4}$ and find, for both cases, high mass loss ($\Delta M/M_{*,i} \sim 1$) and diminished spin angular momentum close to zero at short orbital periods of $\lesssim 10$ days.  At larger orbital periods,  we find negligible mass loss on the order of $2\%$ for the initially highly rotating case and on the order of $0.2\%$ for the initially non-rotating case.  Interestingly, at an initial orbital period of $\sim 20$ days, the final spin angular momentum is comparable between the initial rotation cases.  However, this deviates at longer initial orbital period where the highly spinning case has final spin angular momentum values leveled off at $\sim 2\times 10^{-2} J_{*,i}$ and the non-rotating case decreases to zero as the tidal interaction becomes less significant.   We compare initial stellar rotation results at metallicity $Z=10^{-3}$ and find similar behavior for the $Z=10^{-4}$ case, but with the transition in final mass loss and final spin angular momentum shifted to higher initial orbital periods.  For both initial stellar rotations at metallicity $Z=10^{-3}$, we find high mass loss ($\Delta M/M_{*,i} \sim 1$) and diminished spin angular momentum close to zero orbital periods of $\lesssim 60$ days.  At larger orbital periods,  we find negligible mass loss on the order of $5\%$ for the initially highly rotating case and on the order of $0.1\%$ for the initially non-rotating case.  Interestingly, at an initial orbital period of $\sim 70$ days, the final spin angular momentum  is comparable between the initial rotation cases.  However, this deviates at higher initial orbital period where the highly spinning case has final spin angular momentum values leveled off at $\sim 10^{-2} J_{*,i}$ and the non-rotating case decreases to zero as the tidal interaction becomes less significant.  We compare initial stellar rotation results at metallicity $Z=10^{-2}$ and find overall similar behavior for the $Z=10^{-4}$ and $Z=10^{-3}$ cases, but with the transition in final mass loss and spin shifted to significantly higher initial orbital periods.  This shift is expected because of the strength of the tidal interaction related to the stellar radius at different metallicities, as discussed in Sec.~\ref{subsec:metallicity}.  For both initial stellar rotations at metallicity $Z=10^{-2}$, we find high mass loss ($\Delta M/M_{*,i} \sim 1$) at orbital periods of $\lesssim 3 \times 10^3$.  As noted previously in Sec.~\ref{subsec:metallicity}, for the initially non-rotating case, the tidal interaction at a range of short orbital periods for MS at this metallicity removes mass from the envelope leaving a rotating core with some significant spin.  Overall, the spin angular momentum values are similar for both initial rotations where for shorter initial orbital periods ($\sim 50$ days), the tidal interaction is significant enough although the MS loses considerable mass.  The remnant has stellar spin-up close to the lower threshold for a GRB progenitor (dotted line) at $\sim 5 \times 10^{-5} J_{*,i}$ for the initially non-rotating case and $\lesssim 10^{-5} J_{*,i}$ for the initially highly non-rotating case where it is subjected to more mass loss and subsequently greater diminishment.  At larger orbital periods, there is a regime where the tidal interaction is weaker with significant associated mass loss diminishing the final spin angular momentum.  Finally, at the largest orbital periods, we find negligible mass loss on the order of $\lesssim 10\%$ for both the initially highly rotating and non-rotating cases.  At this transition, the final spin for both initial stellar rotations reaches the GRB threshold limits (green and dotted line) at close to 4 $\times 10^3$ This deviates at higher initial orbital period where the highly spinning case has final spin angular momentum values leveled off at $\sim 10^{-2} J_{*,i}$ and the non-rotating case decreases to zero as the tidal interaction becomes less significant.   In comparing the results at different stellar metallicity, we find that binary interactions with negligible mass loss sustain highly spinning MS above the GRB progenitor threshold given an initially high stellar rotation.
}

\subsection{Additional Tests and Caveats}
There are various mechanisms, such as the Spruit-Tayler dynamo, other magnetic coupling effects, or convection effects \citep{Spruit2002,HWS05, MM14,Ma2019}, that can transport angular momentum out of the star throughout its lifetime. For example, Fig.~3 of \cite{HWS05} shows a pronounced decrease in angular momentum when magnetic fields are present, particularly for the pre-supernova stages of a star's evolution.  However, \cite{Den07} show that some of the assumptions required to invoke these effects are not always valid and, as a result, the loss of angular momentum can be significantly reduced.  \cite{Pot12} analyze the importance of the Spruit-Tayler dynamo in stars over a range of masses, and find that for stars above about 15$M_{\odot}$ it becomes more difficult to sustain the dynamo, and as such the magnetic braking that normally occurs when this dynamo is active may less severe for more MS (particularly in the earlier stages of their evolution). We learned that enabling or disabling this flag, while retaining the default MESA parameters, had no impact on our results. We also investigated the effects of a rotating convection zone within the MS. We began this investigation by setting the  diffusion coefficients as prescribed in \cite{Heger_2000} to their default values according to MESA. We find no change in the final angular momentum of the MS at the end of its lifetime when this flag was turned on.

\section{Discussion}

{ We have investigated the properties of a MS with a closely orbiting BH companion and find a range of parameter space in which the MS terminates with high spin angular momentum.  We conjecture that this then is a plausible model for a highly spinning BH-disk system upon the MS collapse, one that is capable of producing a GRB. In particular, with our simulations of initially non-rotating MS, we show the parameter space which gives rise to the lower limit of final spin angular momentum in Fig.~\ref{fig:main-results2} and Fig.~\ref{fig:main-results3}.  We find a favorable region for relatively short initial orbital periods and low metallicities, where the tidal interaction is stronger and the stellar winds are weaker, to long initial orbital periods and high metallicities, where the tidal interaction is weaker and the stellar winds are stronger.  This region is curtailed once \(R_{*,f}\) approaches \(R_{\rm RL}\) so that RLOF removes angular momentum.  From inspecting the initial conditions associated with these binary parameters, we do not expect tidal synchronization over one orbit.  However, we do find in the cases of relatively short initial orbital periods that synchronization occurs due to the rapid increase in stellar radius over many orbits at the end of stellar lifetime.  Furthermore, because of this lack of tidal synchronization for certain regions of parameter space, we have found cases with initially highly rotating MS that lead to neglible mass loss and sustained spin angular momentum above the threshold for likely GRB progenitor estimates.  For this study, we have chosen very high initial stellar velocity as a rough estimate of an upper limit to the initial rotation.  We note that a more complete analysis of the dependence of final mass loss and spin angular momentum on initial rotation is best performed in a future study coupled with an informed theoretical estimate of the likely initial rotations for MS using population synthesis calculations for these types of binaries.  Furthermore, we have shown MS evolution in binaries within the limits of a one-dimensional, computational domain under the "shellular approximation" in \textsc{MESA}.  A thorough investigation with a three-dimensional code could provide a more robust result of the radial expansion of the MS and detailed evolution of the angular momentum profile.}  We note that after collapse, the BH central engine generally needs a spin parameter (ratio of its angular momentum to maximum possible angular momentum) of at least about 0.5 to launch a jet through the BZ process.  For a $5 M_\odot$ BH this corresponds to roughly $\sim 10^{50}$ g cm$^{2}/s$.  { For our simulations of lower than solar metallicity MS in particular, we find that the star at the end of its life has angular momentum in the range of $\sim 10^{50}$ to $10^{52}$ g cm$^{2}/s$.}

    { Additionally,} we find that modeling the long-term interaction between a MS and BH across stellar lifetime leads to the identification of a parameter region of diminished, final stellar spin angular momentum due to mass loss driven by RLOF as the stellar radius expands prior to collapse.  These results which show a sharp decrease in stellar spin coincident with an increase in mass loss in late-stage stellar evolution.  \citet{Paxton2013} discuss the challenges and uncertainty in modeling the radiation-pressure dominated envelopes in MS at late phases in the evolution with a 1D stellar evolution code.   Nonetheless, we present our results at stellar termination with \textsc{MESA} while noting that follow-up numerical investigations that go beyond its limitations are necessary to resolve these uncertainties.

Regarding the parameter space of encounters with competing effects that change the stellar angular momentum, \cite{Det08} find that the tidal interaction leads to significant spin-up as well as subsequent losses due to mass loss by winds and RLOF between $6-18 M_\odot$ Wolf-Rayet stars at solar metallicity and a $1.4-5 M_\odot$ compact object companions with initial orbital periods less than 24 hours.  Using rapid binary population synthesis, \citet{izzard2004} show the same competition between tidal spin-up and angular-momentum losses from winds and mass transfer in interacting massive-star binaries, and note that the outcome depends on the adopted wind and mass-transfer prescriptions. { Our investigation shows the drop in \(\Delta J_{\rm spin}\) to the left of the boundary of significant mass loss in Fig.~\ref{fig:main-results2} and Fig.~\ref{fig:main-results3}, consistent with the location of the RLOF limit (\(R_{*,f}/R_{\rm RL}=1\)) in Fig.~\ref{fig:main-estimate}.}
 We note that there is additional loss of orbital angular momentum through gravitational radiation \citep{Pet64} for which we have not accounted.  However, for the orbital separations we consider (Newtonian gravity regime), this is a negligible effect and does not significantly affect our results over the timescales we are considering \citep{Map20, Ber20}.  {Other effects (not considered in this study), including common-envelope evolution and dynamical friction, can alter the angular-momentum loss timescale and orbital evolution, shaping the MS’s final angular momentum \citep{izzard2004}.}

 %Other effects (that we do not consider in this study) such as common envelope evolution and dynamical friction can also change the angular momentum loss timescale, significantly affecting the orbital evolution and, in turn, the final angular momentum of the MS. Consistent with this, \citet{izzard2004} show that predicted spins and transient outcomes in massive binaries depend strongly on common-envelope efficiency, wind mass loss, and compact-object formation assumptions

In connecting our progenitor system to a GRB - which results when the highly spinning MS in our system collapses to a BH - we have also not accounted for the potential loss of angular momentum upon collapse of the star. Here, we are making the assumption that a highly spinning MS at the end of its life will produce a highly spinning BH-disk system.  This is reasonably justified as shown in \cite{Jer21}, who demonstrate that highly spinning MS in the ranges we consider end up with a roughly $10 M_{\odot}$ BH remnant, with a spin parameter $a \sim 1$, and with a long-lived disk (all elements to necessary to launch a relativistic GRB jet). 

\cite{Qin18} use MESA to explore a range of conditions, in systems similar to the ones we consider here, where the spin of the second BH is high.  They find that for very closely orbiting binaries, the spin of the second BH (i.e. the BH produced from the MS in the systems we consider) is indeed near maximum ($a \sim 1$), particularly for low metallicity systems where mass loss from line driven winds is relatively small.  \citet{Ma_2023} use MESA to investigate the interaction between a WR star with a BH companion. They find that a WR star at solar metallicity experiences significant angular momentum loss due to stellar winds. In contrast, WR stars with low metallicity retain their angular momentum and are spun up due to tidal interactions. 

Therefore, under the reasonable assumption of a highly spinning MS produces a highly spinning BH-disk system, this central engine with high angular momentum will then produce a longer lasting GRB jet.  This, in turn, can manifest observationally as longer lasting GRB prompt emission, consistent with what is observed in the class of radio bright GRBs.   \\

 We note that in order to make a successful GRB from any system, the polar region (along the spin axis) of the star through which the jet must propagate must be evacuated enough to allow for the GRB jet to punch through the stellar envelope (or, alternatively, the jet must have enough power to allow this to happen) and propagate into the ISM \citep[e.g.][]{MWH01, Pir04, izzard2004, Brom11}. This means we have the additional requirement in these systems that either the MS envelope is stripped before the star collapses, or that during the collapse and jet launch phase, the polar region is sufficiently sparse to allow the launch of a relativistic jet \citep{Burr07}.\footnote{Again, in this paper, we do not model the collapse and jet launch but we refer the reader to \cite{JJN22} who present GRMHD simulations of GRB jets with central engines that have angular momentum similar to what we consider, and show a very powerful magnetically dominated jet is launched, sufficient to punch through a stellar envelope.  See also \cite{Matz03, Brom11, Brom14, LR20, Hutch24, Ham25}.}

We suggest that in some cases (especially for higher metallicity stars with a ``puffier'' density profile or for more vigorous tidal interactions), a denser (relative to a single MS) medium may be formed around the binary system to a larger radius and may potentially lead to a brighter radio afterglow, where the radio flux scales as roughly the square root of the circumbinary density \citep{SPN98}.  The specifics of the mass loss from line driven winds in these systems will also affect the circumbinary density profile. In this paper, we simply offer the suggestion of a varying circumbinary environment depending on the resulting density profiles of star, as seen in Fig.~\ref{fig:radprof}, but leave a detailed investigation of this issue to a following publication.

The tidal interaction between the BH and the MS is the underlying mechanism of the spin-up of our MS, as well as the extended density profile that deviates from the typical $1/r^{2}$ profile of a single MS.  We note that \cite{Sci24} have pointed out that tidal prescriptions in binary stellar evolution codes can sometimes either over or underestimate the strength of the tidal torque and so caution must be taken in over-interpreting results.\\

\section{Conclusions}

  In this paper we have explored the influence of a BH companion on a MS throughout its lifetime, over a range of parameter space varying orbital period, metallicity, initial spin of the star, mass ratio, accretion rate, magnetic fields, and numerical stopping conditions.  Our study is motivated by the interesting result that radio bright GRBs tend to have a longer prompt gamma-ray burst duration compared to radio dark GRBs.  As such, we have explored systems that potentially lead to a MS with more angular momentum (and therefore a longer lived accretion disk and jet which reflects the duration of the prompt emission).  We have focused on a range of parameter space in which the initial orbital period (binary separation) is such that the MS does not initially experience RLOF and tidal interactions still play a significant role in the spin angular momentum evolution of the MS.  We have also briefly discussed the dependence of the density profile of the MS on binary interaction and stellar metallicity (which has implications for the density of the circumbinary medium).

Our main conclusions are as follows:
\begin{itemize}
    \item  { We identify a particular range of parameter space where MS, with closely orbiting BH companions, end their life with high spin angular momentum, conducive to producing a highly spinning BH-disk system that could power a relatively long-lived GRB jet.  } The significant increase in spin angular momentum is due to an appreciable tidal torque in the binary associated with the increase in stellar radius at late-stages of the stellar evolution.  The specific angular momentum of the star at the end of its life aligns with the expectation of the angular momentum needed for a roughly 10 $M_{\odot}$ star to launch a relativistic jet through the Blanford-Znajek process \citep{BZ77}.

    \item  We find consistency between our theoretical estimates and numerical investigations and conclude that the main mechanisms contributing to the final spin  are the tidal interaction and stellar wind.

    \item  { Our numerical investigation of how a BH companion influences the angular momentum of a MS focuses on initial stellar masses 15$M_\odot$ and 25$M_\odot$ with a metallicity range of $10^{-4} - 10^{-2}$ and BH with initial masses 15$M_\odot$ and 10$M_\odot$.  We consider initial orbital periods ranging from short ($\lesssim$ 10 days) to $10^5$ days and evolve the binary throughout stellar lifetime.  
    Our simulations show viable binaries with final MS spin angular momentum above the threshold estimate $\sim 10^{50}-10^{51}$ g$\cdot$cm$^{2}$s$^{-1}$ of a post-collapse black hole mass of 5-10 $M_\odot$ and a black hole spin parameter $\geq 0.5$.  }
    
    \item { Numerical simulations with initially non-rotating MS give the lower limit of final spin angular momentum at stellar termination.  Our results reveal a highly favored parameter space associated with negligible mass loss ($\lesssim 10^{-2}$ of the initial stellar mass) and significant tidal torque at low metallicities ($\sim 10^{-4}-10^{-3}$) and relatively short initial orbital periods ($\sim 20 - 5 \times 10^2$ days) and high stellar metallicities ($\sim 4 \times 10^{-3}-10^{-2}$) and relatively long initial orbital periods ($\sim 2\times 10^3 - 4\times 10^3$ days).  Outside of this region, non-viable binaries are characterized by significant final mass loss close to a factor of unity or at an orbital separation where the tidal interaction is negligible.  Additionally, for the viable progenitor binary systems, we find negligible changes in the orbital separation $(\sim 1\%)$ aside from a decrease $(\lesssim 10\%)$ for a small portion of simulations with negligible mass loss close to the limit of RLOF.  Otherwise, the significant mass loss leads to increases in the orbital separation no greater than a factor of 10.            
    }
    
%    We identify low metallicity ($\sim 10^{-4}-10^{-3}$) MS, within the range of initial masses 15$M_\odot$ and 25$M_\odot$, interacting with both 15$M_\odot$ and 10$M_\odot$ BH for an initial orbital period of $\sim$ 20-500 days as potential progenitor binary systems.  For this range of parameter space, we note significant final stellar spin-up $\sim 10^{-2} J_{*,i} \lesssim 10^{52}$ g$\cdot$cm$^{2}$s$^{-1}$, and negligible final total mass loss on the order of a factor of $\lesssim 10^{-2}$ of the initial mass of the star. At higher metallicities ($\sim 10^{-2}$), these viable progenitor systems disappear as winds drive substantial mass loss and stellar spin-up is suppressed. 
\item { We find for systems with an initially highly rotating massive star, tidal interactions may sustain high spin angular momentum when there is negligible mass loss because the binary is not tidally synchronized.  In these cases, we expect an even wider region associated with GRB progenitors identified by parameters that give rise negligible mass loss for low stellar metallicities ($\sim 10^{-4}-10^{-3}$) and initial orbital periods above $\gtrsim 2 \times 10^2$ days and high stellar metallicities ($\sim 4 \times 10^{-3}-10^{-2}$) and initial orbital periods above $\gtrsim 2 \times 10^3$ days.}

\item  { We find that the range in parameter space of GRB progenitor spin-ups is similar in morphology for all choices of mass ratios where the stellar and BH masses are comparable.  The location of boundary of significant mass loss has the most significant dependence on the chosen MS initial mass rather than BH initial mass.  For all pairs of initial stellar and BH masses, the competing effects of significant tidal torque and mass loss lead to two distinct branches at relatively shorter initial orbital periods $\lesssim 10^3$ days and longer initial orbital periods $> 10^3$ days.  For the first branch, the value of the final stellar spin angular momentum and range in parameter space is greater and more expansive, respectively, for a MS initial mass of $25M_\odot$ more than a MS initial mass of $15M_\odot$.  The parameter region of viable binaries is shifted towards lower metallicity $\lesssim 10^{-3}$ for the MS initial mass of $15M_\odot$.  For this branch, binary systems with a $15M_\odot$ MS are less likely $25M_\odot$ MS than to form 10$M_\odot$ post-collapse BH. 
For the second branch, we have a small region of spin-ups above the lower estimate for a post-collapse BH progenitor at 5$M_\odot$ at high metallicity $\lesssim 10^{-2}$.  Notably, for the binary system of a $15M_\odot$ MS and $15M_\odot$ BH, we have a very high spin-up above the higher estimate that results from the "sweet spot" of maximum tidal torque and minimum mass loss.}

%The most significant dependence on mass ratio is between the lower ($15 M_\odot$) and upper ($25 M_\odot$) limit of stellar mass for the MS, where the final stellar spin-up is greater in magnitude and more expansive in the region of initial orbital period and metallicity for the larger initial stellar mass.}

%\textcolor{red}{
%    \item  We find that the final stellar spin angular momentum is largely insensitive to the initial stellar rotation for the parameter space we have investigated.  For example, in our investigation 
%    at the lowest metallicity ($\sim 10^{-4}$) with an initial orbital period of $\sim 20$ days, tidal interactions efficiently spin up the MS to the GRB progenitor threshold regardless of whether the star begins non-rotating or with substantial initial rotation.  At the highest metallicity ($\sim 10^{-2}$), for the same initial orbital period, winds dominate the evolution and suppress tidal spin-up, again erasing any dependence on the initial rotation state.  For very large orbital periods where the tidal synchronization time scale is larger than the stellar life time, the tidal interactions provide negligible increases in the spin angular momentum and therefore variations in the initial stellar rotation do not contribute significantly.  In all relevant cases, tidal torques act to synchronize the star with the orbit, so the viability of a progenitor system depends primarily on metallicity and orbital configuration rather than on the assumed initial stellar spin.
%    }
\end{itemize}

The results of our work here inform future detailed studies into the physics of gamma-ray bursts.  Specifically, we use the angular momentum of the MS at the end of its life to inform our initial BH and disk angular momentum set-up in upcoming general relativistic magnetohydrodynamic simulations of the GRB jet produced by these systems %(R. Cheng et al. in prep).  
We find specific angular momenta of the MS in our closely orbiting binaries consistent with what others have employed in the literature for their BH-disk simulations  \citep[e.g.][]{JJN22}.\\

Our work has defined a range of parameter space for specific binary systems that could be GRB progenitors, particularly for the sub-class of radio bright GRBs.  Our results can be used to model populations of these systems and estimate their rates (A. Cason et al. in prep), validating whether they align with observed rates of long radio bright GRBs.

\section*{Acknowledgments}

We thank the referee for the detailed and constructive feedback.
We thank the MESA community for helpful conversations and the many resources provided to help run MESA, and Omer Bromberg for interesting discussions on mass transfer in binaries.
This work was supported by the U.~S. Department of Energy through Los Alamos National Laboratory (LANL).  LANL is operated by Triad National Security, LLC, for the National Nuclear Security Administration of U.S. Department of Energy (Contract No. 89233218CNA000001).   Research presented was supported by the Laboratory Directed Research and Development program of LANL project number 20230115ER. We acknowledge LANL Institutional Computing HPC Resources under project w23extremex. Additional research presented in this article was supported by the 
Laboratory Directed Research and Development program of Los Alamos National Laboratory under 
project number 20210808PRD1. LA-UR-24-22983

%%%%%%%%%%%%%%%%%%%%%%%%%%%%%%%%%%%%%%%%%%%%%%%%%%
\section*{Data Availability}

The data underlying this article will be shared on reasonable request
to the corresponding author.

%%%%%%%%%%%%%%%%%%%%%%%%%%%%%%

% BIBLOGRAPHY AND APPENDIX HAVE BEEN COMMENTED 

%%%%%%%%%%%%%%%%%%%%%%%%%

%%%%%%%%%%%%%%%%%%%% REFERENCES %%%%%%%%%%%%%%%%%%

\bibliography{main_refs}{}
\bibliographystyle{aasjournal}

%%%%%%%%%%%%%%%%%%%%%%%%%%%%%%%%%%%%%%%%%%%%%%%%%%

%%%%%%%%%%%%%%%%% APPENDICES %%%%%%%%%%%%%%%%%%%%%
\section{Appendix}

%%%%%%%%%%%%%%%%%%%%%%%%%%%%%%%%%%%%%%%%%%%%%%%%%%%%%%%%%%%%

\subsection{\textsc{MESA} simulations of single star evolution }
\label{sec:appendix_singlestar}

%%%%%%%%%%%%%%%%%%%%%%%%%%%
\begin{figure}
	\includegraphics[width=0.495\columnwidth]{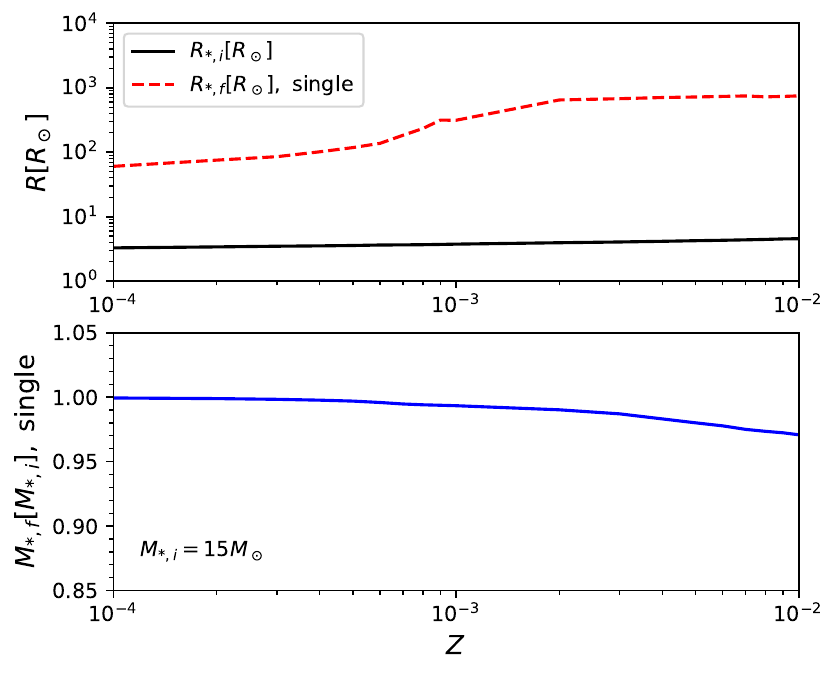}
    	\includegraphics[width=0.495\columnwidth]{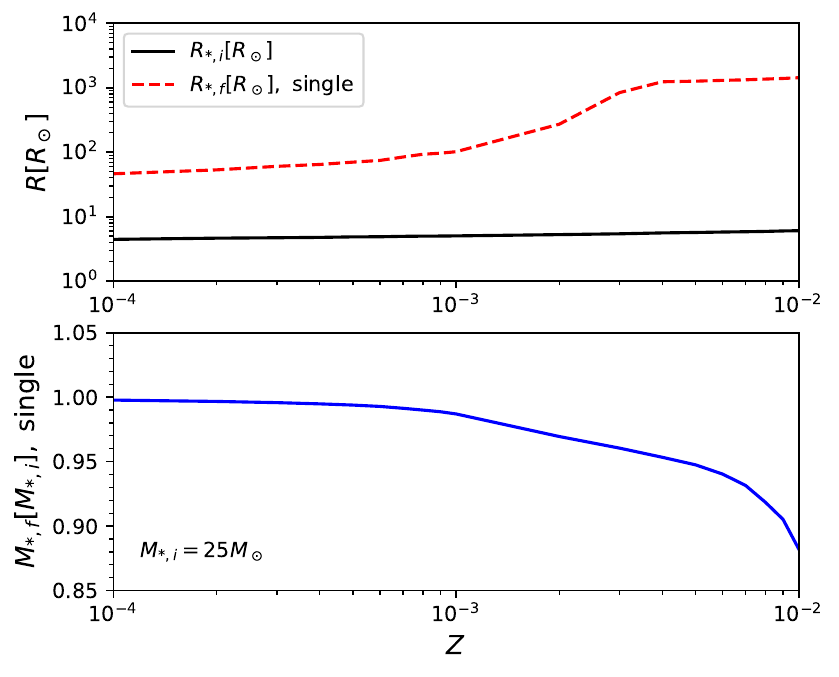}
    \caption{Final stage radius and mass of single MS under a range of metallicities.  We compare results at the end of stellar evolution for MS initially at $M_{*,i} = 15 M_\odot$ (left column) and $25 M_\odot$ (right column) with metallicities between $Z=10^{-4}-10^{-2}$.  In the top panels, we give the initial stellar radii $R_{*,i}$ (black solid line) and the final stellar radii $R_{*,f}$
     (red dashed line), both in units of solar radii.  In the bottom panels, we give the final stellar mass $M_{*,f}$ in units of the initial stellar mass $M_{*,i}$.}
    \label{fig:single_star}
\end{figure}
%%%%

For our \textsc{MESA} simulations in Tab.~\ref{tab:simulation-parameters}, the orbital periods are significantly smaller ($\lesssim 10^5$ days) than the lifetime of the MS ($\sim 6-8$ Myr).  Given the time evolution from ZAMS to the end point prior to collapse, mass loss due to stellar winds may significantly contribute to an increase in stellar radius and subsequently the final stellar spin-up $\Delta J_{\rm spin}$ due to the tidal torque.  Both single star and binary simulations give rise to this rapid increase and its typical evolution is shown in the first row of Fig.~\ref{fig:evol21day}.  With \textsc{MESA}, we simulate single MS evolution prior to collapse and use the final radius $R_{*, f}$ and mass $M_{*, f}$ results in the estimated stellar spin-up in Eq.~\ref{eq:maximum_spin_estimate}.  In Fig.~\ref{fig:single_star}, we show $R_{*, f}$ and $M_{*, f}$ for MS initially at $M_{*,i} = 15 M_\odot$ (left column) and $25 M_\odot$ (right column) with metallicities between $Z=10^{-4}-10^{-2}$.  In the top panels, we compare the initial stellar radii $R_{*,i}$ (black solid line) and the final stellar radii $R_{*,f}$ (red dashed line), both in units of solar radii.  While there is an increase in initial stellar radius from low ($10^{-4}$) to high ($10^{-2}$) metallicity, there is a significant increase of a factor $\sim$ 10 to 100 for the final radius.  In the bottom panels, we show the final stellar mass $M_{*,f}$ in units of the initial stellar mass $M_{*,i}$ and note that mass loss increases with metallicity and at a greater increase for the larger initial mass ($25 M_\odot$).  Although the duration of the rapid increase in radius varies with the mass and metallicity of the MS, from inspection, it may be estimated on average as $\Delta t \sim 10^4$ yr.

%%%%%%%%%%%%%%%%%%%%%%%%%%%%%%%%%%%%%%%%%%%%%%%%%%%%%%%%%%%%

\subsection{Convergence study of \textsc{MESA} simulations}
\label{sec:convergence}

\begin{figure}
	\includegraphics[width=\columnwidth]{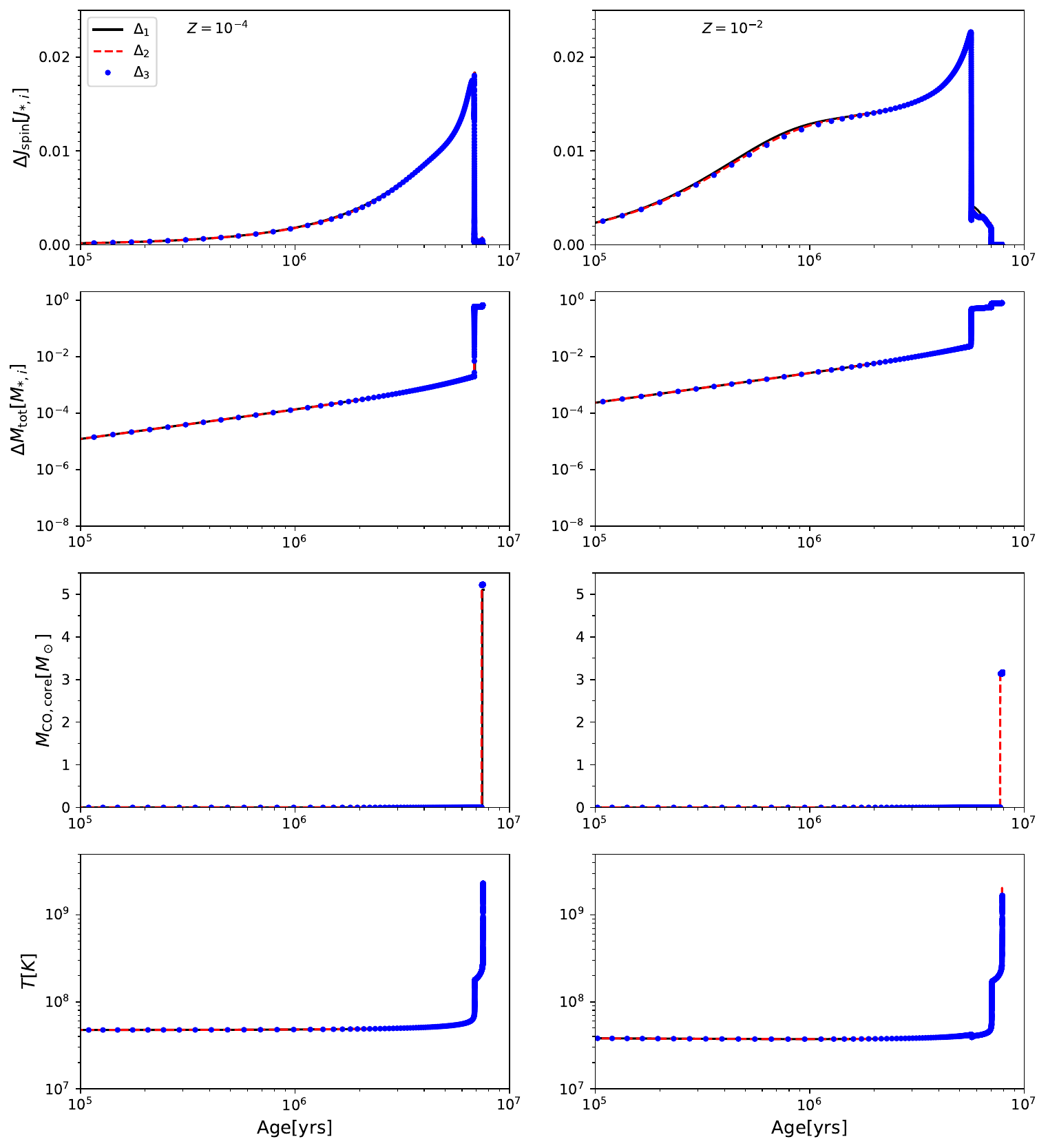}
    \caption{Numerical convergence of stellar spin-up and total mass loss at increasing mass resolution. We compare simulation results for binary models of a $15 M_\odot$ BH and $25 M_\odot$ MS at an initial orbital period of three days
     at three levels of resolution set in {\em MESA} at 1.0e-2 ($\Delta_0$), 1.0e-3 ($\Delta_1$), and 1.0e-4 ($\Delta_2$).  In stellar age, given in years, we show convergence in the change in spin angular momentum $\Delta J_{\rm spin}$, normalized by the initial stellar angular momentum at break-up velocity $J_{*, i}$, and total mass loss $\Delta M_{\rm tot}$, normalized by the initial stellar mass $M_{*, i}$, as well as stellar parameters of carbon-oxygen core mass $M_{\rm CO,\ core}$ and central temperature $T$ in stellar metallicity $Z = 10^{-4}, 10^{-2}$ (left and right panels, respectively).  
     }
    \label{fig:numerical_convergence}
\end{figure}

\begin{figure}
	\includegraphics[width=\columnwidth]{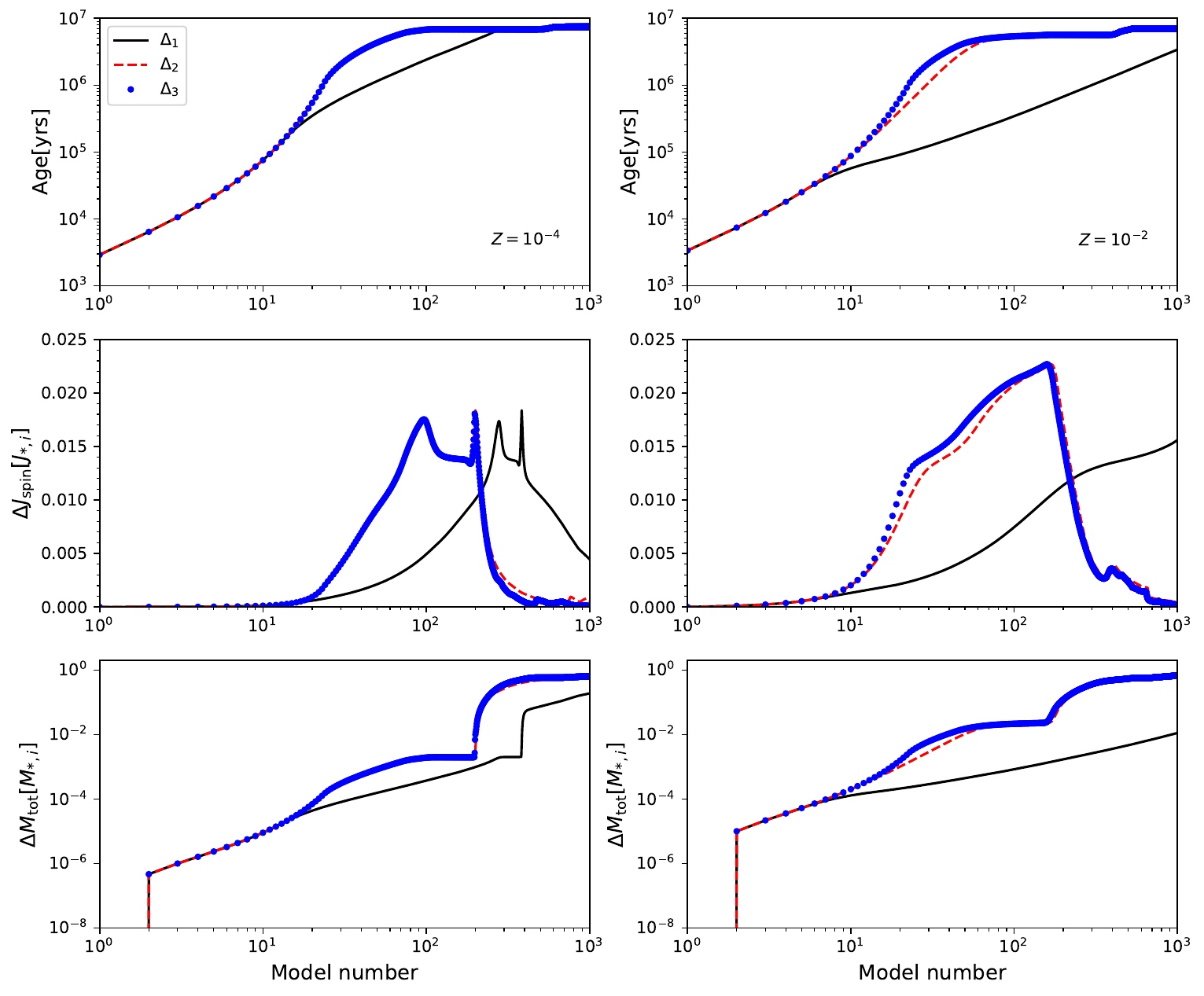}
    \caption
    {
   Evolution in model number of stellar spin-up and total mass loss at increasing mass resolution. We compare simulation results for binary models of a $15 M_\odot$ BH and $25 M_\odot$ MS at an initial orbital period of three days
     at three levels of resolution set in \textsc{MESA} at 1.0e-2 ($\Delta_0$), 1.0e-3 ($\Delta_1$), and 1.0e-4 ($\Delta_2$).  In model number, we show comparably smooth changes for each resolution in stellar age, given in years, and the evolution of the change in spin angular momentum $\Delta J_{\rm spin}$, normalized by the initial stellar angular momentum at break-up velocity $J_{*, i}$, and total mass loss $\Delta M_{\rm tot}$, normalized by the initial stellar mass $M_{*, i}$, in stellar metallicity $Z = 10^{-4}, 10^{-2}$ (left and right panels, respectively). }
    
    \label{fig:numerical_convergence-model-number}
\end{figure}

We have simulated a MS in \textsc{MESA} over its lifetime tracking the evolution of quantities such as angular momentum and mass.  We study the numerical convergence of our simulations where these quantities do not change appreciably over stellar lifetime at increasing mass resolution.  In Fig.~\ref{fig:numerical_convergence}, we show consistent evolution over stellar age, in years, in stellar spin-up $\Delta J_{\rm spin}$, normalized by the initial stellar angular momentum at break-up velocity $J_{*, i}$, and total mass loss $\Delta M_{\rm tot}$, normalized by the initial stellar mass $M_{*, i}$, as well as evolution of carbon-oxygen core mass $M_{\rm CO,\ core}$ and stellar central temperature $T$ for binary models of a $15 M_\odot$ BH and $25 M_\odot$ MS at an initial orbital period of three days at increasing mass resolution with \textsc{MESA} flag {\it max\_dq} at 1.0e-2 ($\Delta_0$), 1.0e-3 ($\Delta_1$), and 1.0e-4 ($\Delta_2$).  We find similar trends for metallicity range $Z = 10^{-2}$ (left panel) and $Z = 10^{-4}$ (right panel).  In Fig.~\ref{fig:numerical_convergence-model-number}, we consider the evolution in model number of stellar spin-up and mass loss for the binary at increasing mass resolution using the models given in Fig.~\ref{fig:numerical_convergence}.  We show comparably smooth changes for each resolution in stellar age, given in years, for the evolution of binary parameters.  For all other simulation results, we use the highest mass resolution, $\Delta_2$.

%%%%%%%%%%%%%%%%%%%%%%%%%%%
%%%%

%%%%%%%%%%%%%%%%%%%%%%%%%%%%%%%%%%%%%%%%%%%%%%%%%%%%%%%%%%

%%%%%%%%%%%%%%%%%%%%%%%%%%%%%%%%%%%%%%%%%%%%%%%%%%%%%%%%%%%

%If you want to present additional material which would interrupt the flow of the main paper,
%it can be placed in an Appendix which appears after the list of references.

%%%%%%%%%%%%%%%%%%%%%%%%%%%%%%%%%%%%%%%%%%%%%%%%%%

%%%%%%%%%%%%%%%

%Extra plots here%%%%

%%%%%%%%%%%%%%%%

% Don't change these lines
\nobreakspace	% typesetting comment

\label{lastpage}
\end{document}